\documentclass[lettersize,journal]{IEEEtran}
\usepackage{amsmath,amsfonts}
\usepackage{mathtools}
\usepackage{algorithmic}
\usepackage{algorithm}
\usepackage{array}
\usepackage[caption=false,font=normalsize,labelfont=sf,textfont=sf]{subfig}
\usepackage{textcomp}
\usepackage{stfloats}
\usepackage{url}
\usepackage{verbatim}
\usepackage{graphicx}
\graphicspath{ {figures/} }
\usepackage{cite}
\usepackage{makecell}
\usepackage{resizegather}
\usepackage{xcolor}

\begin{document}

\title{On Dynamic Ray Tracing and Anticipative Channel Prediction for Dynamic Environments}

\author{Denis Bilibashi, Enrico M. Vitucci, and Vittorio Degli-Esposti

\thanks{This work was funded in part by the Italian Ministry of University and Research (MUR) through the programme ”Dipartimenti di Eccellenza (2018–2022) — Precision Cyberphysical Systems Project (P-CPS),” and in part by the Eu COST Action INTERACT (Intelligence-Enabling Radio Communications for Seamless Inclusive Interactions) under Grant CA20120.}
\thanks{Denis Bilibashi, Enrico~M.~Vitucci, and V.~Degli-Esposti are with the Department of Electrical, Electronic, and Information Engineering "Guglielmo Marconi" (DEI), CNIT, University of Bologna, 40126 Bologna, Italy (e-mail: denis.bilibashi2, enricomaria.vitucci, v.degliesposti @unibo.it)}
}


\maketitle

\begin{abstract}
Ray tracing algorithms, that can simulate multipath radio propagation in presence of geometric obstacles such as buildings, objects or vehicles, are becoming quite popular, due to the increasing availability of digital environment databases and  high-performance computation platforms, such as multi-core computers and cloud computing services. When objects or vehicles are moving, which is the case of industrial or vehicular environments, multiple successive representations of the environment ("snapshots") and multiple ray tracing runs are often necessary, which require a great human effort and a great deal of computation resources.
Recently, the Dynamic Ray Tracing (DRT) approach has been proposed to predict the multipath evolution within a given time lapse on the base of the current multipath geometry, assuming constant speeds and/or accelerations for moving objects, using  analytical extrapolation formulas.  This is done without re-running a full ray tracing for every "snapshot" of the environment, therefore with a great computation time saving. When DRT is embedded in a mobile radio system and used in real-time, ahead-of-time (or anticipative) field prediction is possible that opens the way to interesting applications. In the present work, a full-3D DRT algorithm is presented that allows to account for multiple reflections, edge diffraction and diffuse scattering for the general case where moving objects can translate and rotate. For the purpose of validation, the model is first applied to some ideal cases and then to realistic cases where results are compared with conventional ray tracing simulation and measurements available in the literature.
\end{abstract}


\section{Introduction}
Originally conceived in the seventies and eighties of the last century for optical propagation problems, Ray tracing (RT) algorithms have been developed and used in the last three decades to simulate multipath radio propagation in man-made environments where obstacles, such as buildings, vehicles and machines, can be represented as polyhedrons of a given material \cite{iskander,fuschini2019}. Ray tracing applications have encompassed radio coverage prediction, the simulation of the radio channel, including its multidimensional characteristics, such as time-dispersion and angle-dispersion, which are important for MIMO system design and performance evaluation.
In the last years, as carrier frequencies are extending towards mm-wave and sub-THz frequencies in search for free bandwidth, where ray optics approximations become more valid, RT has become increasingly popular and its performance has improved considerably.

At the same time, public transportation and vehicular applications of wireless systems aimed at improving safety, connectivity, and enabling Connected, Cooperative, Autonomous Mobility (CCAM) technologies are becoming more and more important \cite{Bhat2018}. Due to the high mobility, vehicular channels are characterized by severe time-variability, especially at mm-wave frequencies where Line of Sight (LoS) to Non- Line of Sight (NLoS) transitions become more abrupt due to vanishing diffraction and transmission contributions and Doppler shifts become larger due to the higher carrier frequency \cite{boban1}. Real-time estimation of the channel state, which is crucial to achieve good performance, especially in MIMO and beamforming applications, becomes difficult in dynamic environment. Therefore, new solutions based on the so called "location awareness" are being proposed, including the use of artificial intelligence techniques \cite{Xing2020} or the real-time use of RT within the radio network \cite{iskander,VDE2021}. 

Unfortunately, the application of traditional RT models to vehicular environments, or dynamic environments in general, requires the solution of complex problems. Due to the high variability, realistic, spatially-consistent simulation of propagation in such cases requires to consider a high number of successive environment configurations (or “snapshots”): for each one of them, a new RT simulation needs to be carried out from scratch, which can result in an overwhelming computation time, not to mention the problems related to handling a huge number of environment description files and output files. The availability of fast ray tracing algorithms is therefore critical to future wireless applications in dynamic environments. 

Methods to predict the evolution of radio visibility prior to the actual ray tracing run, based on geometric techniques \cite{he2019,hussain2019}, methods to interpolate the evolution of the channel’s coefficients based on the speed of the moving transmitter (TX) and receiver (RX) \cite{nuckelt2015} as well as fast techniques to estimate Doppler profiles \cite{azpilicueta} have been proposed.   
Only recently, a new paradigm where RT is redesigned specifically for dynamic environments, such as vehicular and smart-factory environments, has been proposed using the definition Dynamic Ray Tracing (DRT) \cite{bilibashi,bilibashi2,qua}. The DRT concept implies the use of a dynamic environment database, where all the moving objects, including the radio terminals, vehicles and machine’s moving parts, are described in terms of their position, geometric shape, and their speed and acceleration at a given reference time instant $t_0$. Based on such a description and on a traditional RT run for the environment configuration at $t_0$, DRT predicts the dynamic evolution of the multipath geometry using an analytic extrapolation technique for any successive instant in time $t_{0}+\Delta t$, as long as $\Delta t < T_C$, where $T_C$ is the multipath coherence time (or lifetime), i.e. the time during which no major path should either appear or disappear.  Such a technique allows analytical field prediction for any number of snapshots within $T_C$ using a single RT run, with a reduction of the computation time of several orders of magnitude. Of course, the estimation of $T_C$ is a key point that deserves investigation: a discussion on $T_C$ can be found in the results section of this paper. 

Differently to previous work, a complete description of moving objects as solid bodies is considered in DRT, which allows the correct sliding of reflection and diffraction points on the bodys' surfaces to be taken into account. Therefore a high degree of accuracy can be achieved even when obstacles and radio terminals are relatively close to each other, which is common in both vehicular and industrial environments. Note that, if DRT is used in real-time and instantaneous speeds and accelerations for all environment’s moving parts are available for $t=t_0$, then even anticipative prediction for $t>t_0$ is possible within $T_C$: this is a very attractive possibility to enable a timely selection of the proper channel coding and/or beam for transmission and realize truly dependable wireless communications in dynamic environments.  When prediction for $\Delta t > T_C$ is needed, a conventional RT run must be carried out again based on the new geometric environment configuration that might be predicted using kinematics theory \cite{Martin1968,taylor2005}.

In the present work, the DRT concept is described and developed into a full-3D approach that includes multiple-bounce reflection, diffraction and diffuse scattering. With respect to the work presented in \cite{qua} the method is fully extended to edge diffraction and the rotation of solid bodies is also taken into account to achieve a general approach for dynamic environments. The DRT concept is described in more detail in section II, the algorithm formulation is presented in section III and applied to both ideal cases and real-world cases where it is checked against literature results including measured power-delay profiles. Using comparison with measurements and different choices of the maximum extrapolation time, the multipath coherence time $T_C$ is estimated and the possibility of anticipative channel prediction is presented.

\section{The DRT Concept}
\begin{figure}[h!]
	\centering
	\includegraphics[width=3in]{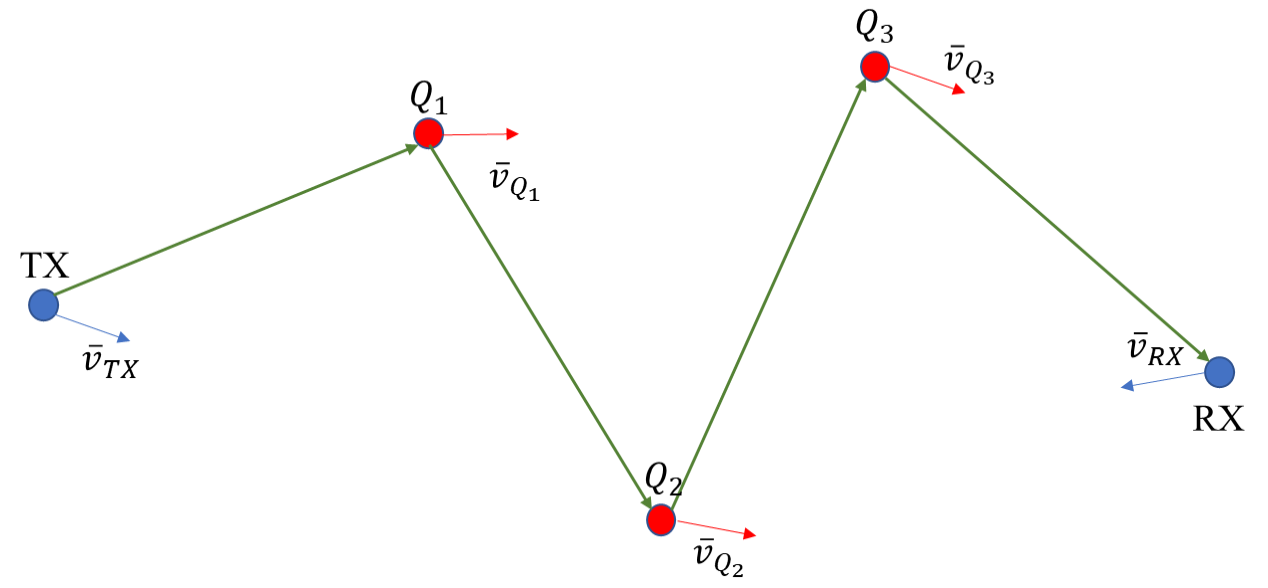}
	\caption{Representation of single multi-bounce ray path.}
	\label{ray_path}
\end{figure}

A single multi-bounce ray path can be represented as a polygonal chain where TX, the interaction points (e.g. reflection points or diffraction points) and RX are defined as the vertexes of the chain (see Fig. \ref{ray_path}). The evolution of this chain in a dynamic environment can be described only if we know how its vertexes will move in time. While the motion of the terminals is independent, the motion of the interaction points depends on the motion of the terminals and of objects generating these interaction points as well as on their instantaneous positions. Furthermore, the motion of the interaction points is not constant even if the terminals motion is constant. \par

The computation of the dynamic evolution of paths can of course be accomplished by multiple RT runs on multiple environment descriptions corresponding to successive "snapshots" of the environment in time. Nonetheless, this technique requires the creation of a very large number of environment databases as well as an equally large number of RT runs making it impractical if an accurate description is required and moreover it will require a huge CPU time. \par
The proposed DRT algorithm is based on the combination of a classical 3D image-based RT approach with an analytical extrapolation of multipath evolution using a proper description of the dynamic environment \cite{bilibashi}, \cite{bilibashi2}. A dynamic environment database is introduced that describes the geometry of the environment also including the motion characteristics of both the radio terminals and the rigid bodies that can move, such as vehicles or machines. The database is divided into two parts. The first one provides a geometrical description of the environment for a given time instant $t_0$. Objects are described here - as in most RT models - as polyhedrons with flat surfaces and right edges, although an extension to curved surfaces is possible. In the second part, the dynamic parameters of every terminal and object are provided. In particular, the objects are modelled as "rigid bodies" with a roto-translational motion, that can be described in a complete way by providing translation velocity, rotation axis, angular velocity and the corresponding accelerations. \par
%
%
A traditional RT prediction is carried out for the initial time $t_0$, then the positions of all interaction points are extracted. From the initial positions, the motion of each vertex of the chain within $T_C$ must be computed. Our assumption is that the complete motion of the radio terminals is a-priori known. If this is not true, we assume at least that the the terminals' instant velocity and acceleration at $t_0$ is known, so that the positions $\overline{r}_P(t)$ of TX or RX on a time instant $t=t_{0}+\Delta t$ can be calculated using the Taylor series formula:
\begin{equation}
\begin{gathered}
 \overline{r}_P(t_{0}+\Delta t) = \overline{r}_P(t_{0}) + \overline{v}_{P}(t_0) \Delta t +  \\
 \frac{1}{2} \overline{a}_{P}(t_0) \Delta t^2 + O(\Delta t^3)
\end{gathered}
\label{Taylor}
\end{equation}
where $\overline{v}_{P}(t_0)$ and $\overline{a}_{P}(t_0)$ are the velocity and acceleration of P, respectively, at $t_0$, and the symbol $O(\Delta t^3)$ means that all the variations higher than $2^{nd}$ order, i.e. variations of accelerations, are neglected.

Now the problem is that the interaction points' positions, velocities and accelerations is unknown and needs to be determined for every time instant $t_{0}+\Delta t$ on the base of the current positions, velocities and accelerations of Tx, Rx and of the obstacles generating such interactions. For this purpose, we developed closed form formulas as explained in Section III.

The analytical extrapolation procedure is valid as long as the multipath structure remains the same, i.e., no path disappears and no new path appears to significantly change such structure. This time interval will be referred to in the following as \textit{multipath~lifetime}~$(T_C)$. As long as $\Delta t < T_C$ , DRT allows to extrapolate the multipath - and therefore the total field, the time-variant channel's transfer function, Doppler's shift frequencies, time delays, etc. - for every time instant $t_{0}+\Delta t$ without re-running the RT engine, and therefore at only a fraction of the computation time. In fact, the computationally most expensive part of a RT algorithm consists in determining the geometry of the rays, which requires checking the visibility and obstructions between all objects in the database, in order to establish the existence of each of the traced rays. With the DRT approach, this is done only once at time $t_0$, while in the subsequent time instants within $T_C$ we rely on analytical prolongation of the same rays, which is computationally much faster.

Also, we believe that it is a reasonable assumption to neglect the temporal variations of acceleration during the time $T_C$, in accordance with eq. (\ref{Taylor}): in fact, in vehicular scenarios the multipath structure usually varies on a faster time scale than acceleration, as discussed in Section IV.

Since DRT naturally accounts for speeds and accelerations, another advantage is that dynamic channel parameters, such as Doppler information, are derived in closed form, without resorting to finite-difference computation. A more detailed explanation about Doppler frequency calculation is described in Appendix A.

\section{DRT Algorithm Description}
In this Section, the DRT algorithm formulation is described for single and multiple bounce rays, including specular reflection and edge diffraction, based on the outcome of a single RT simulation at the initial time $t_0$, and on the knowledge of the dynamic parameters of TX/RX, and of the objects in the propagation environment. \par

\subsection{Reflection Points' Calculation}
In this subsection, the DRT algorithm is initially explained for the single reflection case. Starting from this, further extensions of DRT to double reflection and multiple reflections, are presented in the following subsection. 

\subsubsection{Single Reflection Procedure}
A simple case is considered with a reflecting wall laying on a plane $\Pi^I$, and transmitter (TX) and receiver (RX) located at different distances from $\Pi^I$. As a start, we consider the case where TX and RX move with instantaneous speeds $\overline{v}_{TX}$ and $\overline{v}_{RX}$, while the reflecting wall is at rest. Without loss of generality, we can assume a proper reference system so that the wall lies on the plane of equation $y=0$, while TX and RX lie on the xy plane. Then, by using the image method and through simple geometric considerations we can derive the reflection point position ($Q_R$) at the considered time instant, which is located at the intersection between the reflecting plane and the line passing through RX and the image of TX:

\begin{equation}
\begin{gathered}
\begin{cases}
x_{Q_{R}}(t) = x_{TX}(t) + \frac{x_{RX}(t)-x_{TX}(t)}{y_{RX}(t)+y_{TX}(t)}~y_{TX}(t) \\
y_{Q_{R}}(t) = 0 \\
z_{Q_{R}}(t) = z_{TX}(t) + \frac{z_{RX}(t)-z_{TX}(t)}{y_{RX}(t)+y_{TX}(t)}~y_{TX}(t)
\end{cases}
\end{gathered}
\label{ref_point}
\end{equation}
When TX/RX move with a certain speed, the corresponding reflection point "slides" along the plane surface: therefore, by deriving $Q_R$ coordinates with respect to time, the instantaneous velocity of the reflection point ($\overline{v}_{Q_{R}}=v_{Q_{R,x}}\hat{x}+v_{Q_{R,z}}\hat{z}$) can be determined, by using the derivative chain rule. 

For example, the x-component of $\overline{v}_{Q_{R}}$ is:
\begin{equation}
\begin{gathered}
v_{Q_{R,x}} = \frac{\partial x_{Q_R}}{\partial t} = \frac{\partial x_{Q_R}}{\partial x_{TX}}~v_{TX,x} + \frac{\partial x_{Q_R}}{\partial y_{TX}}~v_{TX,y}  \\ + \frac{\partial x_{Q_R}}{\partial x_{RX}}~v_{RX,x} + \frac{\partial x_{Q_R}}{\partial y_{RX}}~v_{RX,y}.
\end{gathered}
\label{ref_point_velo}
\end{equation}
The closed-form expressions of the derivatives in eq. (\ref{ref_point_velo}) are reported in Appendix B.

This approach is similar to the one adopted in \cite{qua}, but in the following it will be extended to a more general case, where the reflecting wall has a roto-translational motion, and TX, RX, and the reflecting wall can vary their velocities during time, so their motion is accelerated.
The basic idea is to use a local reference system integral with the wall (local frame) so that we can resort to the previous case where the wall is at rest and apply equations (\ref{ref_point}) and (\ref{ref_point_velo}) again, as shown below. \par

Moreover, in the present work we make a complete characterization of the reflection point motion, including acceleration: in fact, except very particular cases, $Q_R$ velocity is not constant in time, i.e. the reflection point has an accelerated motion. Therefore, with a similar method as in (\ref{ref_point_velo}), the acceleration of the reflection point ($\overline{a}_{Q_{R}}=a_{Q_{R,x}}\hat{x}+a_{Q_{R,z}}\hat{z}$) can be calculated by deriving $\overline{v}_{Q_{R}}$ with respect to time. For instance, the x-component of the acceleration will be:  
\begin{equation}
a_{Q_{R,x}} = \frac{\partial v_{Q_{R,x}}}{\partial t} = \frac{\partial}{\partial t} \frac{\partial x_{Q_{R}}}{\partial t} 
\label{ref_point_acc}
\end{equation}
The complete expression of $\overline{a}_{Q_{R}}$, and the detailed computation of $\overline{v}_{Q_{R}}$ and $\overline{a}_{Q_{R}}$ are presented in Appendix B. \par 

In Fig. \ref{scenario_before_after}, an example of a single reflection scenario is presented. For the sake of simplicity, we refer to a bi-dimensional case where TX/RX are located in a horizontal plane $z=z_0$ and moving with arbitrary speeds, meanwhile the reflecting wall is located along a vertical plane $y=y_0$, and is then represented with a straight line. Moreover, the wall rotates around a vertical axis, and the intersection of the rotation axis with the plane $z=z_0$ provides the rotation center $O^I$. The local reference system $O^Ix^Iy^Iz^I$ centered in $O^I$ (local frame), is also represented in the figure. The axes of the local frame are oriented so that at each time instant the reflecting wall is laying on the $z^Ix^I$ plane of equation $y=0$, in order to apply eq. (\ref{ref_point}).
The scenario represented in the figure is simplified for ease of readability, but the presented method is general, so the TX/RX positions and velocities, the wall plane and the rotation axis can be arbitrarily oriented with respect to the global reference system $Oxyz$.

\begin{figure}[!ht]
	\centering
	\includegraphics[width=3.5in]{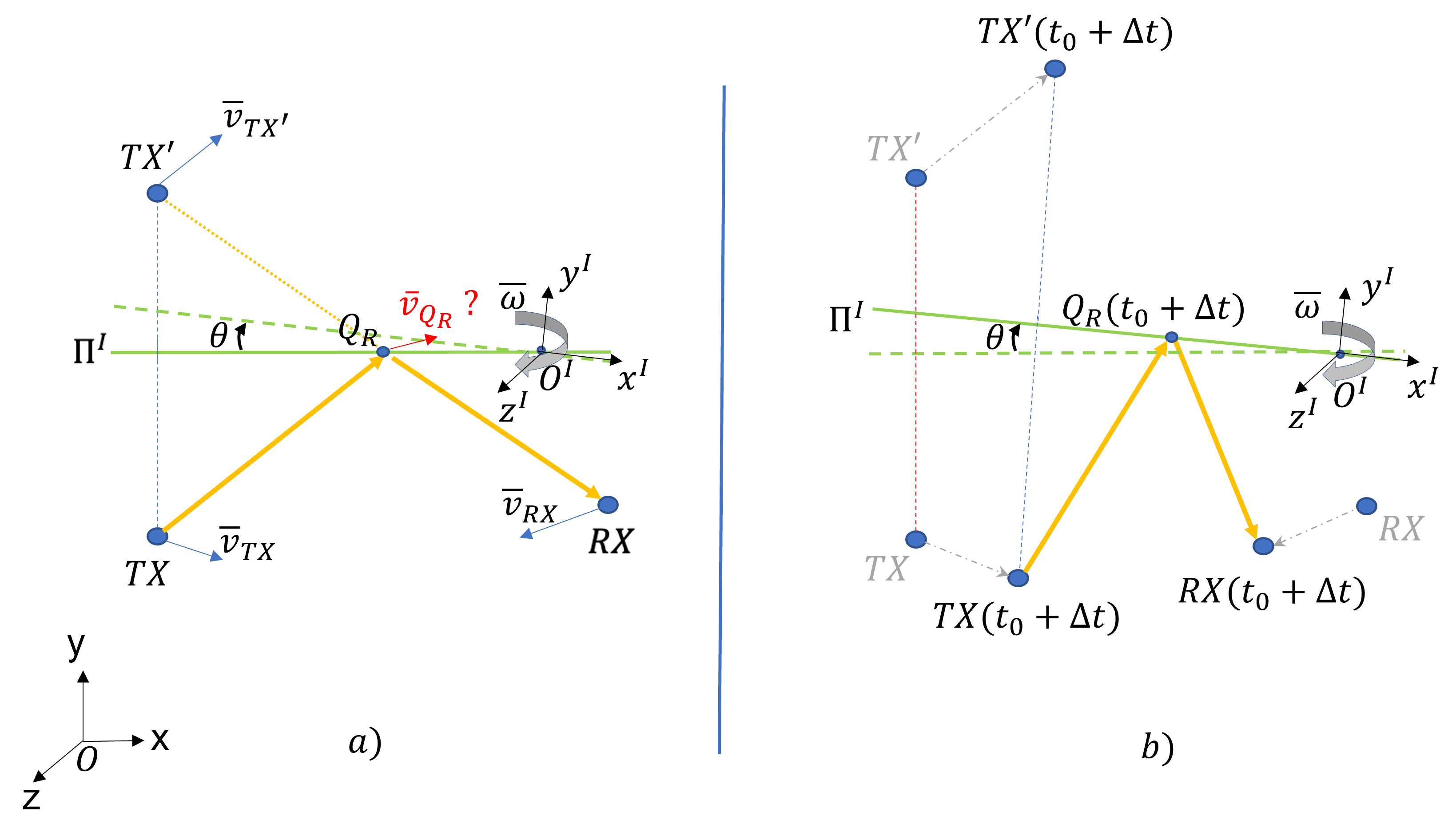}
	\caption{Example of a single reflection scenario with the plane $\Pi^I$ that rotates w.r.t. to the global frame. a) Scenario at $t=t_0$ and instantaneous velocities of TX, image-TX, RX, and reflection point $Q_R$. b) Scenario at $t=t_0+\Delta t$.}
	\label{scenario_before_after}
\end{figure}

Fig. \ref{scenario_before_after}a) shows the configuration of the system at a certain time $t=t_0$, when the wall plane $\Pi^{I}$ is parallel to the plane $xz$ of the global frame $Oxyz$. The instantaneous velocities of TX ($\overline{v}_{TX}$), image-TX ($\overline{v}_{TX'}$), reflection point ($\overline{v}_{Q_R}$) and RX ($\overline{v}_{RX}$) at $t=t_0$ are depicted as well. The instantaneous angular velocity of the wall plane is expressed by the vector $\overline{\omega}=\omega \hat{k}$, where $\omega=\frac{d \alpha}{dt}$ is the scalar angular velocity, and $\hat{k}$ is a unit vector parallel to the rotation axis, and properly oriented according to the right-hand rule. In the example of Fig. \ref{scenario_before_after}, we have $\hat{k}=-\hat{z}^I$, as the wall is rotating clockwise around the z-axis of the local frame.

Fig. \ref{scenario_before_after}b) shows the configuration of the system in a subsequent time instant $t=t_0+\Delta t$, when the wall plane has rotated clockwise by an angle $\theta$, and TX/RX have moved to different positions: the result of this motion is a shift of the reflection point $Q_R$ along the reflecting wall.  

In Fig. \ref{scenario_before_after} a wall rotating around a fixed rotation axis is shown, but in general, a translational motion of the wall plane can be also present, in addition to the rotational motion.
In the general case of roto-translational motion, any point Q of the wall will have a different speed, given by \cite{Martin1968}:
\begin{equation}
\overline{v}_Q=\overline{v}_{\Pi^I}+\overline{\omega} \times \overline{O^IQ}
\label{RigidBodyMotion}
\end{equation}
where the symbol "$\times$" stands for the cross vector product, $\overline{v}_{\Pi^I}$ is the translation velocity, common to all the points of the plane, and $\overline{O^IQ}$ is the position vector of the considered point Q w.r.t. the origin $O^I$ of the local frame, positioned on the (instantaneous) rotation axis.
\par

The first step of the DRT procedure consists in the computation of the (instantaneous) position of the reflecting point $Q_R$. Thanks to the adoption of the local frame, this can be accomplished through eq. (\ref{ref_point}), but in order to do that, we need to transform the coordinates of TX and RX into the local frame associated with the wall. To this end, we observe that the positions of TX/RX with respect to the global and local frames are related each other through the following \textit{coordinate transformation} \cite{Martin1968}:
\begin{equation}
\begin{gathered}
    \overline{r}_{TX}^0(t) = \Bar{\Bar{R}}(t) \cdot \overline{r}_{TX}^{I}(t) + \overline{r}_{0^I}(t) \\
    \overline{r}_{RX}^0(t) = \Bar{\Bar{R}}(t) \cdot \overline{r}_{RX}^{I}(t) + \overline{r}_{0^I}(t)
\end{gathered}
\label{coordinate_transf}
\end{equation}
with
\begin{center}
\begin{small}
$\overline{r}_{TX}^0(t)=[x_{TX}(t)~y_{TX}(t)~z_{TX}(t)]^T$ \\
$\overline{r}_{RX}^0(t)=[x_{RX}(t)~y_{RX}(t)~z_{RX}(t)]^T$
\end{small}
\end{center} 
and 
\begin{center}
\begin{small}
$\overline{r}_{TX}^I(t)=[x_{TX}^I(t)~y_{TX}^I(t)~z_{TX}^I(t)]^T$ \\
$\overline{r}_{RX}^I(t)=[x_{RX}^I(t)~y_{RX}^I(t)~z_{RX}^I(t)]^T$ 
\end{small} 
\end{center}
being the position vectors of TX/RX w.r.t. the global and local frame, respectively, where $\Bar{\Bar{R}}(t)$ is the (instantaneous) rotation matrix and $\overline{r}_{0^I}(t)=[x_{O^I}(t)~y_{O^I}(t)~z_{O^I}(t)]^T$ is the position vector associated with the origin point $O^{I}$ of the local frame. By inverting eq. (\ref{coordinate_transf}), we can determine the positions $\overline{r}_{TX}^I$, $\overline{r}_{RX}^I$ of TX and RX w.r.t. to the local frame, and then apply eq. (\ref{ref_point}) to find the coordinates of $Q_R$.

In the simple example of Fig. \ref{scenario_before_after} (clockwise rotation around the z-axis), the rotation matrix at the time instant $t=t_0+\Delta t$ is given by:
\begin{equation*}
\Bar{\Bar{R}}(t_0+\Delta t)=
\begin{pmatrix}
cos\theta & sin\theta & 0\\
-sin\theta & cos\theta & 0\\
0 & 0 & 1
\end{pmatrix}
\end{equation*}
In a generic time instant $t=t_0+\Delta t$ within the multipath lifetime $T_C$, the instantaneous rotation angle $\theta(t)$ can be obtained in the following way, similarly to eq.(\ref{Taylor}) and neglecting time variations of angular acceleration:
\begin{equation*}
\theta(t)=\theta(t_0)+\omega\Delta t+\frac{1}{2}\frac{d\omega}{dt}\Delta t^2
\end{equation*}
Of course, \textit{coordinate transformation} must be applied to the components of $\overline{v}_{TX}$, $\overline{v}_{RX}$ and $\overline{a}_{TX}$, $\overline{a}_{RX}$ as well.

The next step of the DRT procedure consists in the computation of the velocity and acceleration of $Q_R$ which can be obtained through eq. (\ref{ref_point_velo}) and (\ref{ref_point_acc}). In fact, one of the main advantages of the DRT approach is that we can derive $\overline{v}_{Q_R}$ and $\overline{a}_{Q_R}$ analytically, without need to resort to time differences methods. However, projecting the components of $\overline{v}_{TX}$, $\overline{v}_{RX}$ and $\overline{a}_{TX}$, $\overline{a}_{RX}$ on the local frame is not enough to apply eq. (\ref{ref_point_velo}) and (\ref{ref_point_acc}), which have been obtained from eq. (\ref{ref_point}), i.e. assuming that the reflecting wall is at rest.  Since now the reflecting wall is moving, and then the local frame is in motion w.r.t. the global reference system, the velocities and accelerations of TX/RX must be preliminary transformed according to the \textit{relative motion transformations}, to obtain "relative" velocities and accelerations w.r.t. an observer located in the origin of the local frame \cite{taylor2005}. For instance, the velocity and acceleration of TX must be transformed in the following way: 
\begin{gather}
\overline{v}_{TX}^{I} = \overline{v}_{TX}^{0} - \overline{v}_{\Pi^{I}} - \overline{\omega} \times \overline{r}_{TX}^{I} \label{relative_vel} \\
\overline{a}_{TX}^{I} = \overline{a}_{TX}^{0} - \overline{a}_{\Pi^{I}} - \dot{\overline{\omega}} \times \overline{r}_{TX}^{I} - 2 \overline{\omega} \times \overline{v}_{TX}^{I} \label{relative_acc} \\ \nonumber
-\overline{\omega} \times (\overline{\omega} \times \overline{r}_{TX}^{I})
\end{gather}
where the headers $"I"$ and $"0"$ are associated with the velocities and accelerations seen by an observer located in the origin of the local and global frame, respectively, $\overline{r}_{TX}^{I}=\overline{r}_{TX}^{0}-\overline{r}_{O^{I}}$ is the position vector of TX in the local frame, $\overline{v}_{\Pi^{I}}$, $\overline{\omega}$ are the translation and angular velocities of the wall plane $\Pi^{I}$, and $\overline{a}_{\Pi^{I}}$, $\dot{\overline{\omega}}=\frac{d\overline{\omega}}{dt}$ are the corresponding translation and angular accelerations, respectively.
The terms $\dot{\overline{\omega}} \times \overline{r}_{TX}^{I}$, $2 \overline{\omega} \times \overline{v}_{TX}^{I}$ and $\overline{\omega} \times (\overline{\omega} \times \overline{r}_{TX}^{I})$ in eq. (\ref{relative_acc}) are also known as Euler's, Coriolis', and centrifugal acceleration, respectively.

It is worth noting that the accelerations $\overline{a}_{\Pi^{I}}$, $\dot{\overline{\omega}}$ are assumed to be constant in the time interval $[t_0~t_0+T_C]$, while $\overline{v}_{\Pi^{I}}$, $\overline{\omega}$ in eq. (\ref{relative_vel}) are \textit{instantaneous} velocities, computed as:
\begin{equation*}
\begin{gathered}
\overline{v}_{\Pi^{I}}(t)=\overline{v}_{\Pi^{I}}(t_0)+\overline{a}_{\Pi^{I}}\Delta t \\
\overline{\omega}(t)=\overline{\omega}(t_0)+\dot{\overline{\omega}} \Delta t
\end{gathered}
\end{equation*}

In practice, with the \textit{relative motion transformations} (\ref{relative_vel}) and (\ref{relative_acc}) we turn the original problem into an equivalent problem, where the wall plane is at rest and the TX/RX velocities and accelerations are modified, according to the point of view of an observer located in the origin of the local frame. It is worth noting that, even in the simple case of TX/RX moving with constant speed, TX and RX are accelerated in the equivalent problem: this acceleration is caused by the angular rotation ${\overline{\omega}}$ of the wall plane, according to eq. (\ref{relative_vel}) and (\ref{relative_acc}).
\par

Once the reflection point position, velocity and acceleration have been determined in the local frame using eq. (\ref{ref_point}), (\ref{ref_point_velo}), and (\ref{ref_point_acc}), the following inverse transformations need to be applied: 
\begin{itemize}
	\item back-transformation to get $Q_R$ coordinates, and $\overline{v}_{Q_{R}}$ and $\overline{a}_{Q_{R}}$ components w.r.t. the global reference system (inverse \textit{coordinate  transformation}).
	\item back-transformation to get the velocity and acceleration ($\overline{v}_{Q_{R}}$ and $\overline{a}_{Q_{R}}$) relative to an observer located in the origin of the global reference system (inverse \textit{relative motion transformation}):
	\begin{gather}
	    \overline{v}_{Q_R}^{0} = \overline{v}_{Q_R}^{I} + \overline{v}_{\Pi^{I}} + \overline{\omega} \times \overline{r}_{Q_R}^{I} \label{global_vel_ref_point} \\
        \overline{a}_{Q_R}^{0} = \overline{a}_{Q_R}^{I} + \overline{a}_{\Pi^{I}} + \dot{\overline{\omega}} \times \overline{r}_{Q_R}^{I} + 2 \overline{\omega} \times \overline{v}_{Q_R}^{I} \label{global_acc_ref_point} \\ \nonumber
        +\overline{\omega} \times (\overline{\omega} \times \overline{r}_{Q_R}^{I})   
	\end{gather}
\end{itemize}

\subsubsection{Generalization to Multiple Reflections}
In the case of multiple reflections, the motion of a certain reflection point is influenced by the motion of the previous or latter reflection points.  
However, the method discussed above can be extended in a straightforward way to a multiple-bounce case. 

Let's consider for simplicity a double reflection case: instead of using TX, we can resort to the use of the image-TX ($TX^{'}$) with respect to the first wall plane, and after that we can analyze the reflection on the second wall: in practice, we replace $TX$ with $TX^{'}$ and we bring the computation back to the single-reflection scenario analyzed in the previous section. This means that in a case with two reflecting walls, $TX^{'}$ and $RX$ are used to compute the motion of the reflection point along the second reflecting wall ($Q_{R2}$), by applying the same procedure as for the single-bounce scenario. Then, once $Q_{R2}$ is known, it is used as a new virtual receiver to compute, in addition to the TX location, the position and the motion of the reflection point along the first reflecting wall ($Q_{R1}$). 

This approach can be iterated in a similar way for the case of more than 2 reflections. In practice, we compute the image-TX ($TX^{'}$), the image of the image ($TX^{''}$), etc., until we reach the last reflecting wall, then a "back-tracking" procedure is used: we start from RX and the last reflecting wall, we apply all the equations of the previous section to compute the last reflection point, and then we move back towards TX, to trace the motion of all the remaining reflection points.  

In order to apply all the equations of the previous section for the determination of the reflection points motion, we need to apply all the necessary transformations from the global to the local frame, and vice-versa, for each of the reflecting walls.

Moreover, a preliminary computation is needed to compute the position and the instantaneous velocity for each of the image-transmitters. This can be done in a straightforward way by relying on the local frame, and using the image principle. For example, the image-TX with respect to the first wall ($TX^{'}$), will have the same $x^I$ and $z^I$ coordinates as TX, and opposite $y^I$ coordinate. Similarly, the velocity components will be:

\begin{equation}
\begin{gathered}
\begin{cases}
v_{TX^{'},x}^{I} = v_{TX,x}^{I} \\
v_{TX^{'},y}^{I} = -v_{TX,y}^{I} \\
v_{TX^{'},z}^{I} = v_{TX,z}^{I}
\end{cases}
\end{gathered} 
\end{equation}

Once $\overline{v}_{TX^{'}}$ is computed in the local frame, it can be expressed w.r.t. to an observer located in the origin of the global frame, according to the relative motion transformation:
\begin{equation}
\overline{v}_{TX^{'}}^{0}=\overline{v}_{TX^{'}}^{I}+\overline{\omega} \times \overline{r}_{TX^{'}}^{I}
\end{equation}
In a similar way, the acceleration of the image-TX, $\overline{a}_{TX^{'}}$, can be also found.
The DRT algorithm then proceeds according to steps described above, to find the coordinates, velocities and accelerations for each of the reflection points.
\par

\begin{figure}[h!]
	\centering
	\includegraphics[width=2.5in]{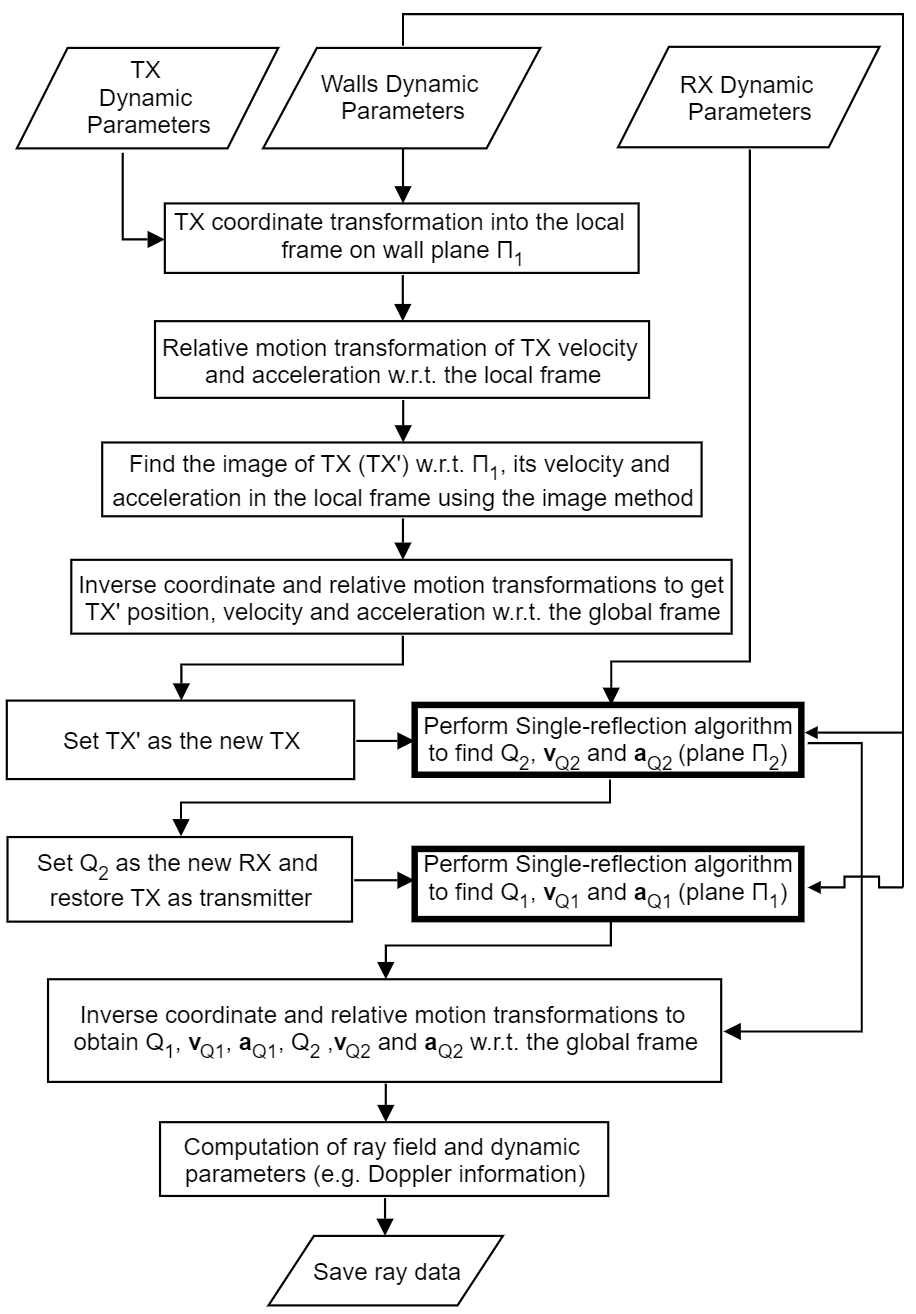}
	\caption{Flowchart showing the DRT algorithm for a double reflection case.}
	\label{flowchart}
\end{figure}

The whole DRT algorithm is summarized by the flowchart depicted in Fig. \ref{flowchart}, with reference to a double-bounce case, for the sake of simplicity. 

Once the geometric part is done, i.e. the analytical prolongation of all the rays in the considered time instants within $T_C$ has been completed, the very last step of DRT consists in the re-computation of the field associated to each ray. This is done in a straightforward way as the geometry of the rays is known, by applying the Fresnel's reflection coefficients and the ray divergence factor, as usually done in standard RT algorithms \cite{fuschini2015,vitucci2019}. It is worth noting however, that ray's field computation is based on analytical formulas and is therefore orders of magnitude faster than ray's geometry computation \cite{fuschini2015}.

\subsection{Diffraction Points' Calculation}
The DRT algorithm can be further extended to diffraction, modeled with a ray-based approach according the Uniform Theory of Diffraction (UTD) \cite{UTD}.
A method to track the motion of diffraction points in an analytical way, as previously done for the reflection points, is outlined in this section. We present only the single-diffraction case for the sake of brevity, but the procedure can be extended in a straightforward way to multiple diffractions, as well as to combinations of multiple reflections and diffractions.

\begin{figure}[!ht]
	\centering
	\includegraphics[width=3.2in]{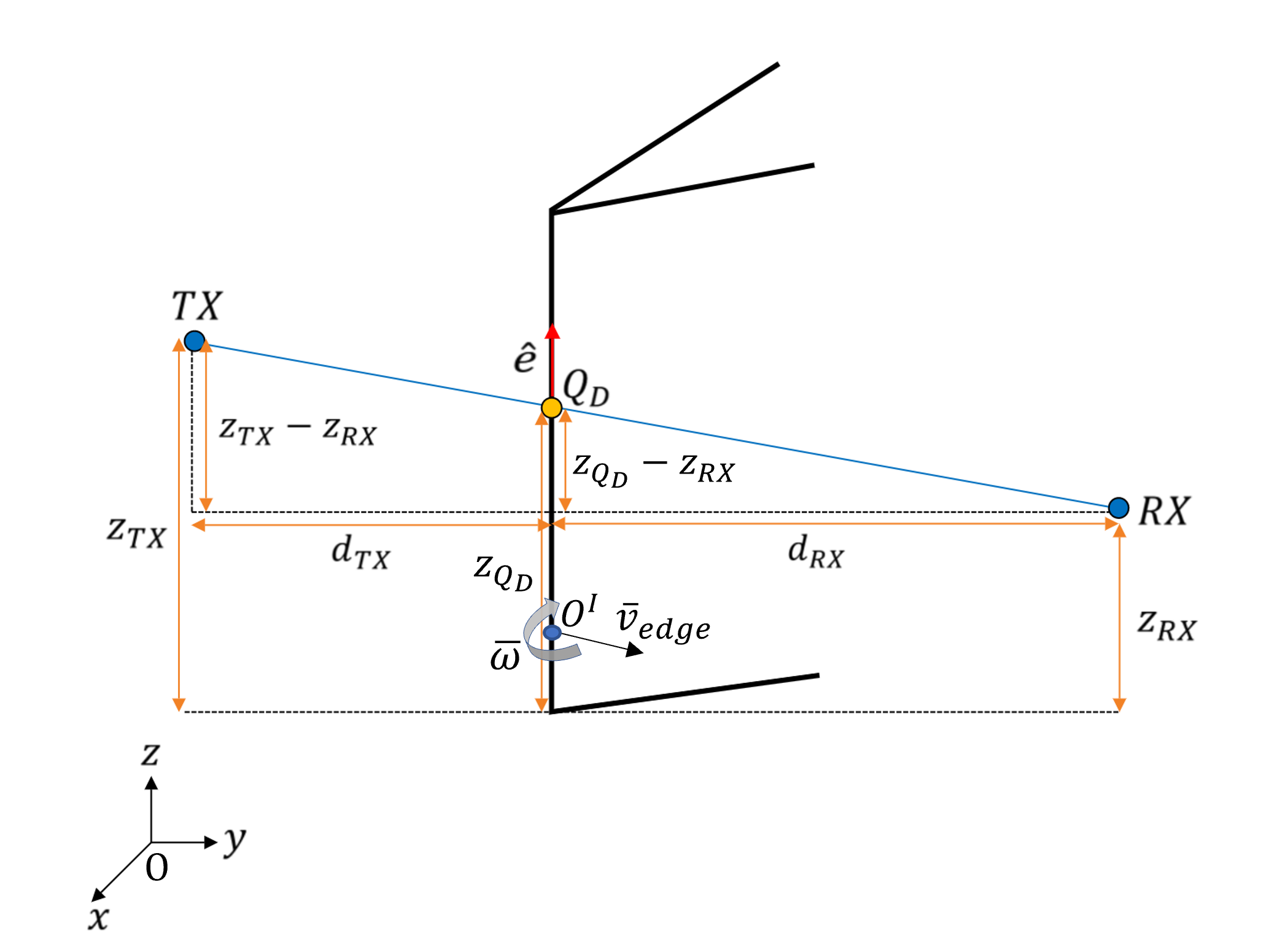}
	\caption{Example of edge diffraction, and related geometry for the computation of the diffraction point.}
	\label{diffraction}
\end{figure}

In Fig. \ref{diffraction} diffraction from an edge formed by two adjacent walls is illustrated, where the unit edge vector $\hat{e}$ is chosen to be aligned with the $z$-axis of the reference system $Oxyz$, with no loss of generality. For the sake of simplicity and with no limitation - as the diffracted rays lay on the Keller's cone and share the same geometric properties \cite{Keller} - we represent in Fig. \ref{diffraction} an "unfolded" diffracted ray, i.e. the diffraction plane has been rotated to be coincident with the incidence plane. 

Since the diffraction point ($Q_D$) is constrained to move along the edge, the $x$ and $y$ coordinates of $Q_D$ are known, then only $z_{Q_D}$ remains to be computed: this can be done in a simple way by using the similar triangles properties.  \par
Looking at Fig. \ref{diffraction}, we see that two similar triangles are formed. The sides of these triangles are proportional each other, so the following relation holds: 
\begin{equation}
z_{TX}-z_{RX} : d_{TX}+d_{RX} = z_{Q_D}-z_{RX} : d_{RX}
\end{equation} 
where 
\begin{equation}
\begin{gathered}
d_{TX}(t) = \sqrt{(x_{TX}(t)-x_{Q_D})^2+(y_{TX}(t)-y_{Q_D})^2} \\[1ex]
d_{RX}(t) = \sqrt{(x_{RX}(t)-x_{Q_D})^2+(y_{RX}(t)-y_{Q_D})^2}
\end{gathered}
\label{dist2D_TXRX}
\end{equation}
are the 2D distances of TX and RX from the edge, respectively. Hence, the z-coordinate of $Q_D$ is given by: 
\begin{equation}
z_{Q_D}(t) = z_{RX}(t) + \frac{d_{RX}(t)\cdot[z_{TX}(t)-z_{RX}(t)]}{d_{TX}(t)+d_{RX}(t)}
\label{diff_point}
\end{equation}
The instantaneous velocity of $Q_D$ ($\overline{v}_{Q_D}=v_{Q_{D,z}}\hat{z}$) can be computed by time deriving $z_{Q_D}$. The detailed calculation of $v_{Q_{D,z}}$ is presented in Appendix C.\par
Similarly, by further deriving $\overline{v}_{Q_D}$, we can find the acceleration of the diffraction point, $\overline{a}_{Q_D}=a_{Q_{D,z}}\hat{z}$, not reported here for the sake of brevity.\par
The expression in (\ref{diff_point}) and the related velocity and acceleration $\overline{v}_{Q_D}$, $\overline{a}_{Q_D}$ are valid and do not require any further computation in case the terminals are moving but the edge is at rest. However, in general an edge might be part of a moving object and then might be moving with a certain roto-translational velocity. In particular, we assume that the edge has a rotational motion with instantaneous angular velocity $\overline{\omega}$, and is also translating according to the instantaneous velocity $\overline{v}_{edge}$. In the example of Fig. \ref{diffraction}, the edge is rotating clockwise around the x-axis. 

As for reflections, we can compute the instantaneous position and the motion of the diffraction point if we assume a proper local reference $O^Ix^Iy^Iz^I$, with the origin located in the rotation center $O^I$, and the z-axis parallel to the edge. By doing so, the same procedure used for reflections, can be followed to transform velocities and accelerations, and finally find $z_{Q_D}$ as well as $v_{Q_{D,z}}$ and $a_{Q_{D,z}}$.  \par 
The final step of DRT is, as usual, the computation of the updated UTD coefficients, and then, of the total diffracted field, at the considered time instant.

\subsection{Diffuse scattering}
Diffuse scattering is modeled according to the Effective Roughness approach \cite{vitucci2019}, which is based on a subdivision of each surface into tiles and on the application of a virtual scattering source to the centroid of each tile. Therefore, the  calculation of scattering points' position and speed is straightforward, as it boils down to the application of basic kinematic equations for the motion of rigid bodies' surface points. For instance, equation (\ref{RigidBodyMotion}) can be used to compute each scattering point's velocity if the rototranslation speed of the body is known.

\section{Results}
The presented algorithm is validated over two main case studies. In the first case the results are obtained in a simple vehicular environment. In the second case, a realistic scenario is considered, and results are compared with measurements carried out in an intersection in the city of Lund, Sweden \cite{abbas}. 
\subsection{Case Study 1: Ideal Street Canyon}

\begin{figure}
    \captionsetup[subfigure]{font=scriptsize,labelfont=scriptsize}
    \centering
    \subfloat[\label{bus_middle}]{\includegraphics[width=1.4in]{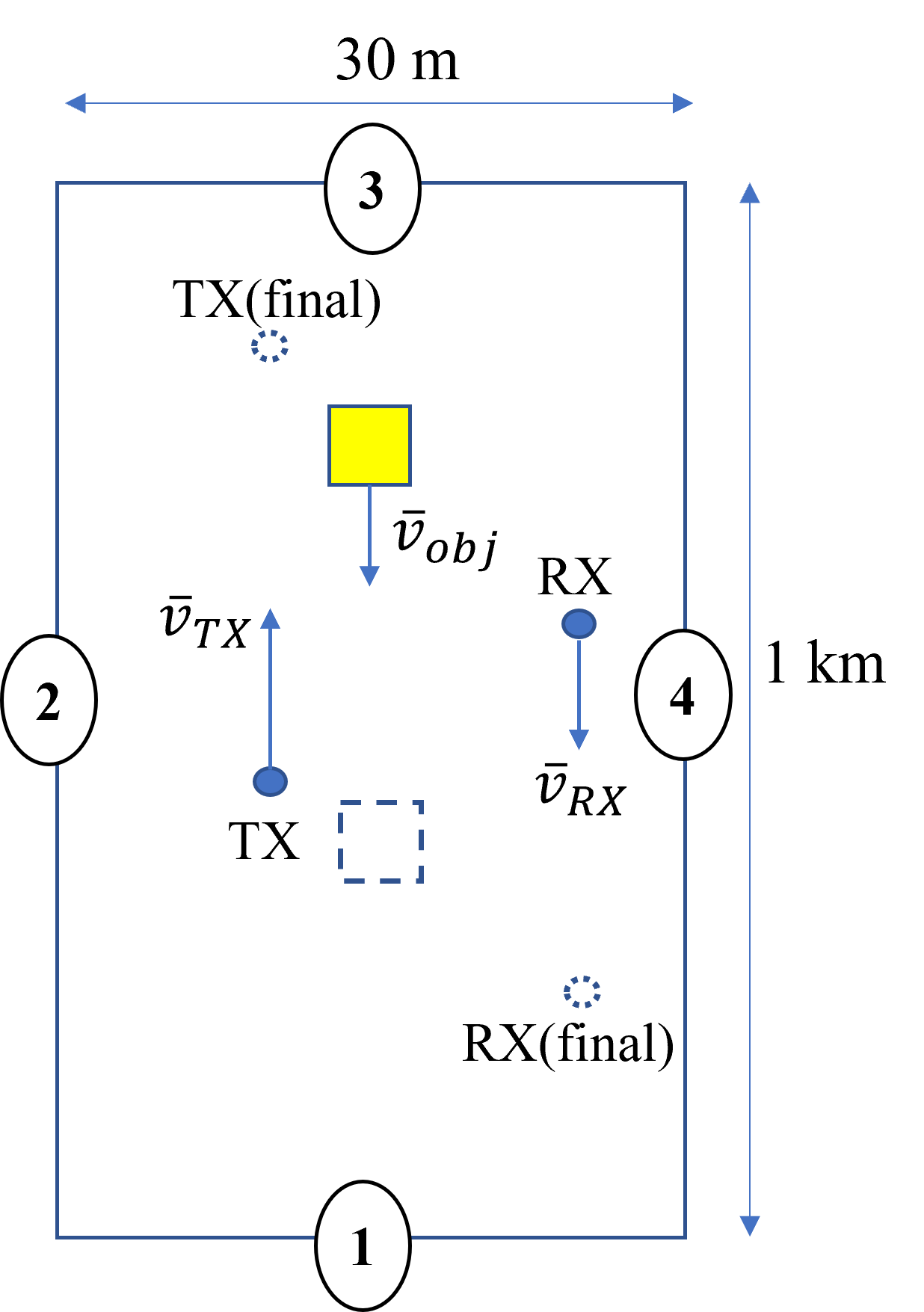}}
    \quad
    \subfloat[\label{bus_aside}]{\includegraphics[width=1.41in]{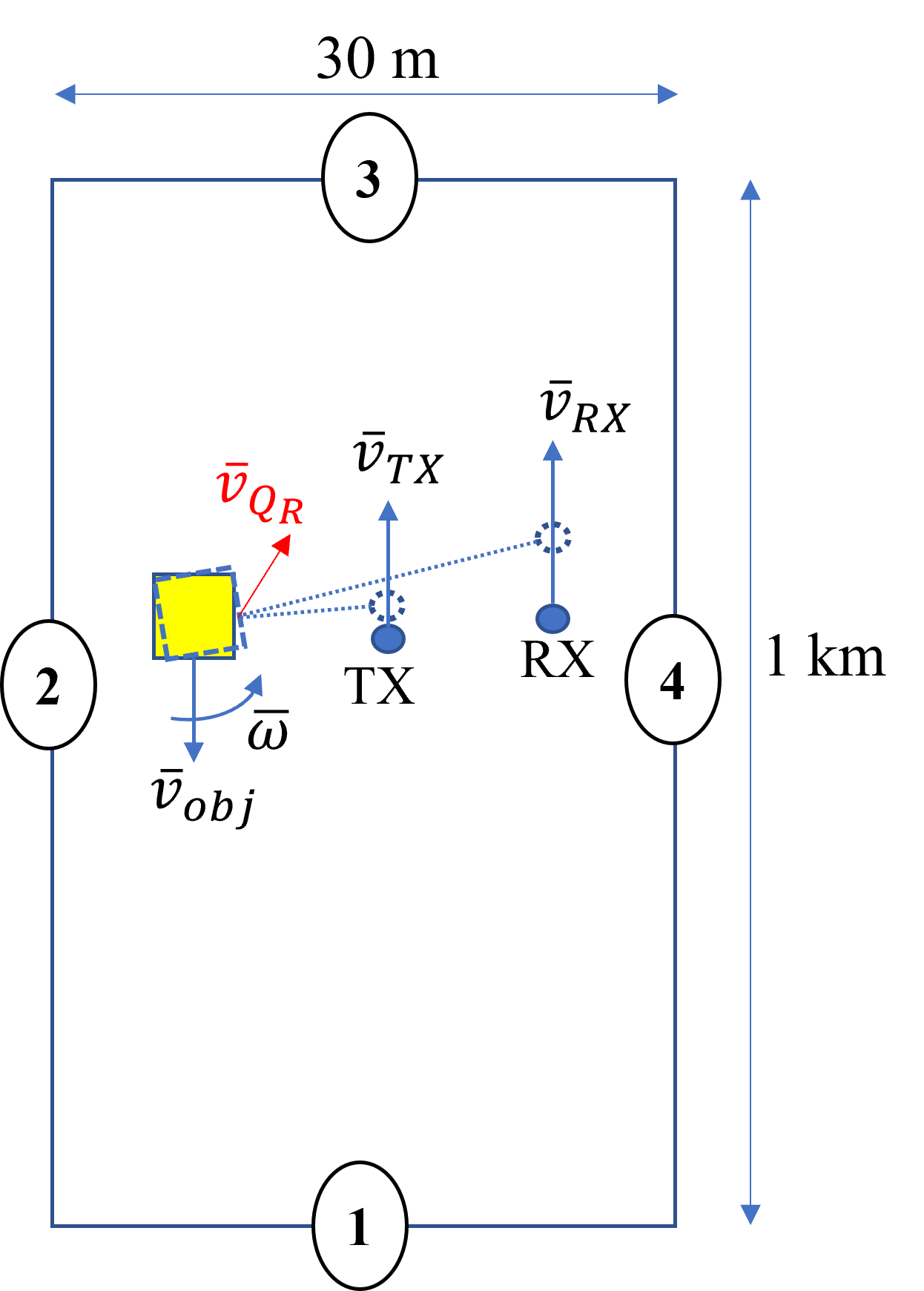}}
    \caption{Ideal street canyon with a moving parallelepiped representing a bus and two moving terminals (TX and RX, not to scale).  The initial positions of the terminals and the bus are depicted with solid lines,  the final positions are shown with dashed lines. The bus is driving in the middle lane in (a) and in the side lane in (b).}
    \label{street_canyon}
\end{figure}

In this section the results are obtained using the DRT approach in a vehicular environment composed of a street canyon, a moving parallelepiped made of metal representing a bus, and two moving radio terminals. The propagation environment consists of an ideal street canyon 1km long and 30m large, with building walls on both sides and no intersections. Two reflecting walls are also present at its end sections. 
We first consider a vehicle-to-vehicle (V2V) communication scenario, where the two vehicles carrying TX and RX terminals drive in opposite lanes and the bus is moving on a middle lane between the two terminals as shown in Fig. \ref{bus_middle}. Therefore, the bus can generate a reflection from its front. We assume that TX is moving toward the end of the street at a constant speed of 50 km/h, RX and the bus are moving in the opposite direction at constant speeds of 36 km/h and 30 km/h, respectively. In the simulations, we considered reflections up to the second order and constant velocities for terminals and moving object. \par
To perform the simulation of the above-mentioned scenario, omnidirectional antennas were chosen with a transmit power of 1 W operating at a carrier frequency of 3 GHz. The antennas were placed on top of the TX and RX terminals at a height of 1.75m.  

\subsubsection{Power Doppler Profile}
\begin{figure}[h!]
	\centering
	\includegraphics[width=3.4in]{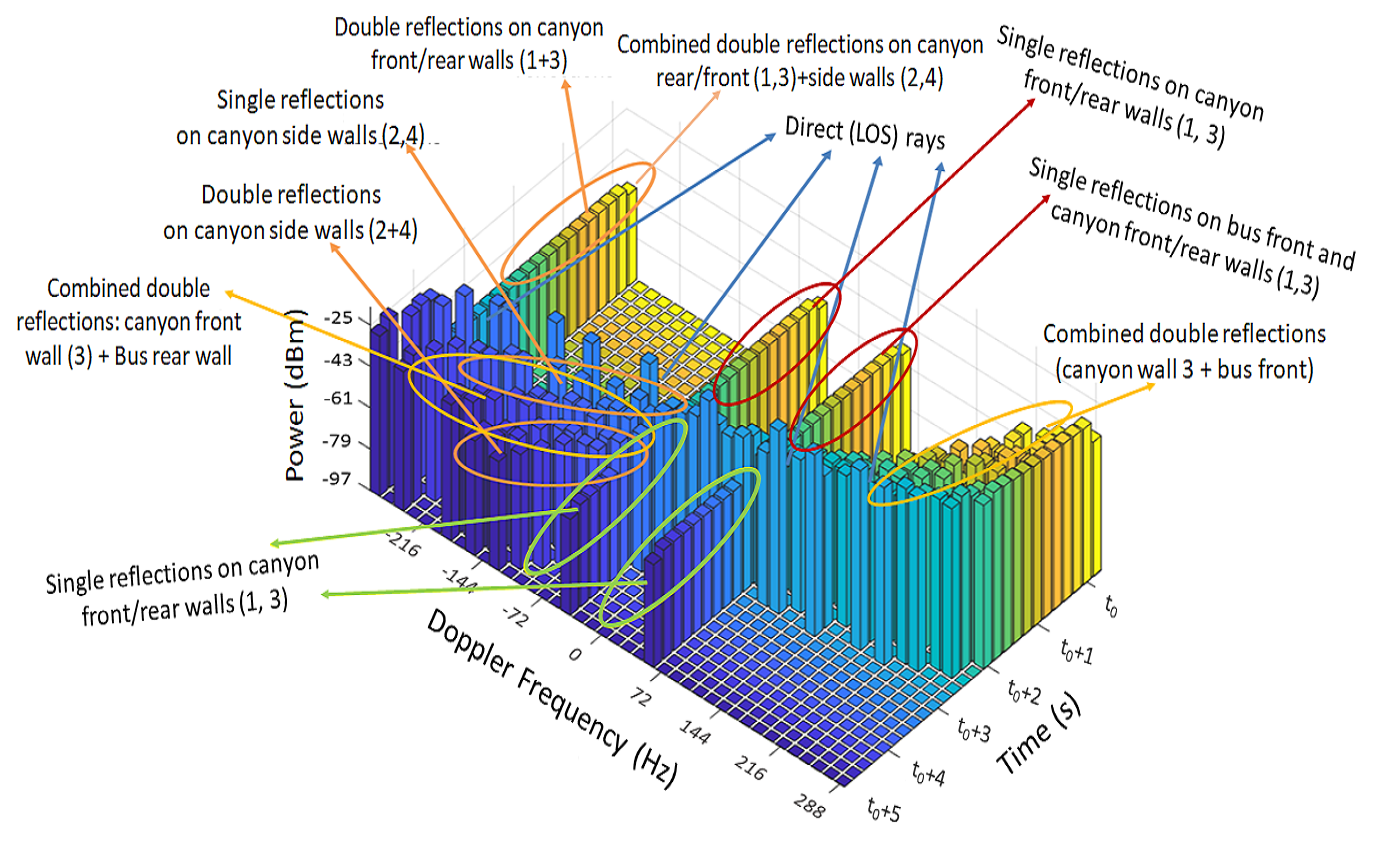}
	\caption{PDfP evolution in a V2V scenario with a simulation time of 5 seconds.}
	\label{PDfP}
\end{figure}

Figure \ref{PDfP} shows the evolution of Power-Doppler frequency profile (PDfP) obtained through DRT with a total simulation time of 5 seconds and discretized into "bins" for display purposes, with a time step of 200 ms and a Doppler frequency step of 14.34 Hz. The Doppler shift is computed for each ray using eq. (\ref{doppler_freq}) (see Appendix A). For each bin, an incoherent sum of the power contribution of each ray whose time of arrival and Doppler shift fall within the bin is performed. 

Since the scenario is simple and the obstructions or abrupt changes are minimal, we perform DRT simulation with only two $T_C$ with length 3 s and 2 s. The reason to use two $T_C$ is related to the contributions from the parallelepiped representing a bus: since the bus passes through the LoS line in the middle of the considered time span, the reflection from the front of the bus disappears and therefore we must consider the first $T_C$ expired and refresh the multipath structure with a new RT simulation.\par 
The other main contributions to PDfP (LoS, single and double reflections) are tagged in the plot and are present mainly throughout the whole simulation period. The most interesting trend in the contributions can be seen in the direct ray which from a positive Doppler shift in the first part of the simulation becomes negative with an inflection when TX and RX passes by each other. The same trend can be seen even for the reflections from the side walls but with lower values and less abrupt transitions. 

\subsubsection{Absolute value of error}
\begin{figure}[h!]
	\centering
	\includegraphics[width=3in]{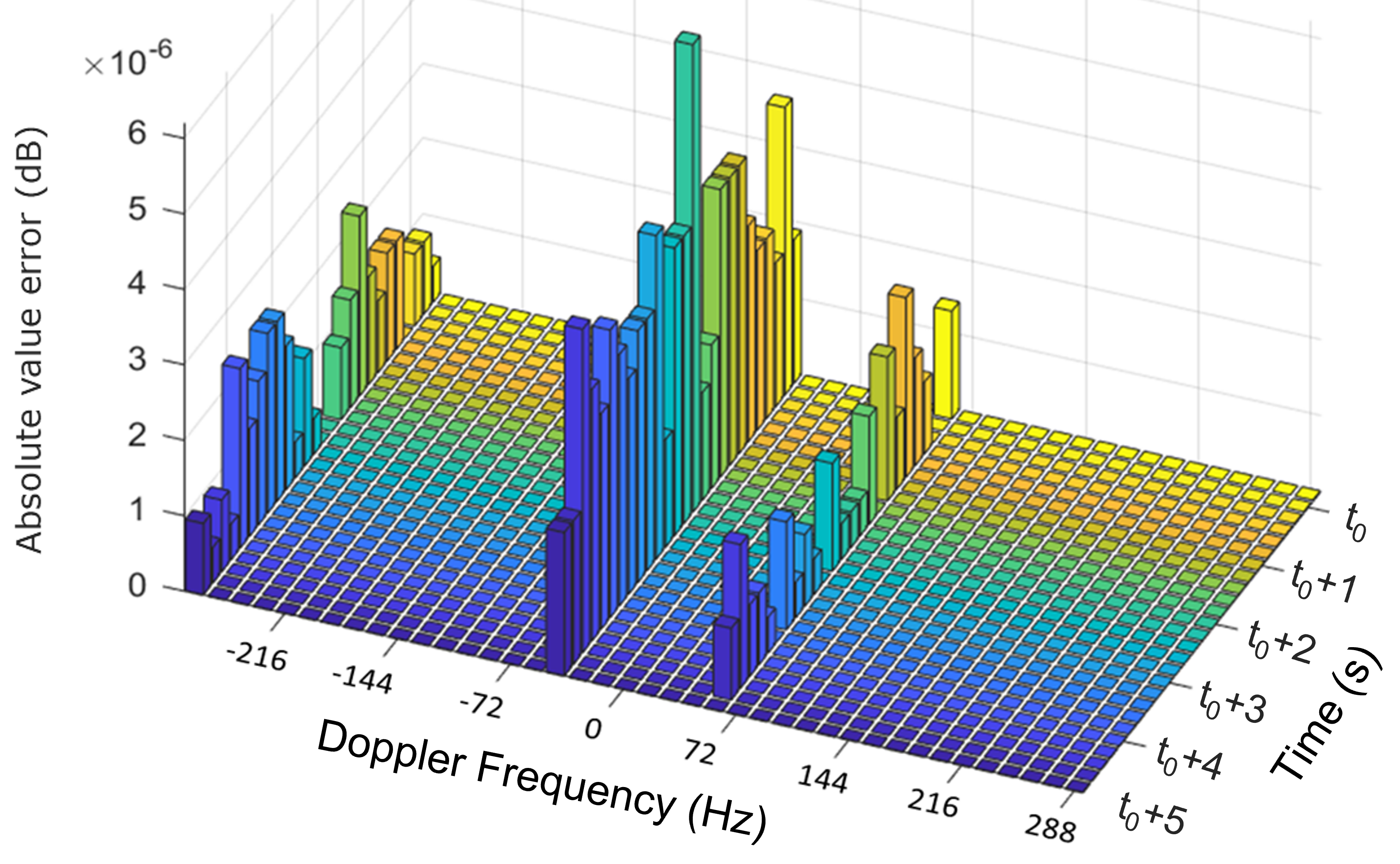}
	\caption{Absolute value of error between standard multiple-run RT and DRT analytical approach}
	\label{absolute_error}
\end{figure}

We have also compared the DRT analytical approach with the classical approach based on "snapshots", where the RT simulation is repeated at every discrete time instant. In Fig. \ref{absolute_error}, we show the absolute value of the error of the estimated power by DRT w.r.t RT in each of the time-Doppler bins of fig \ref{PDfP}. The absolute error is computed as:
\begin{equation*}
e_{ij}^{ABS}(dB)=|P_{ij}^{DRT}(dBm)-P_{ij}^{RT}(dBm)|
\end{equation*}
where $i,j$ are the indices associated to the time and Doppler bins, respectively.
Fig. \ref{absolute_error} shows that the error is virtually zero, as even in the worst case its value is several order of magnitudes lower than 1 dB. The non-zero values in Fig. \ref{absolute_error} are essentially caused by numerical inaccuracies in floating-point operations, and propagation of the errors when applying several successive reference systems transformations.
However, it is interesting to observe that the error becomes more evident in cases where there is a strong acceleration of the reflection point, e.g. when the reflecting wall is perpendicular to motion direction of the vehicles. This acceleration becomes more relevant when TX gets nearer to the wall, and also when the wall is part of a moving object. 

\subsubsection{Bus moving on a side lane and then rotating}
\begin{figure}[h!]
	\centering
	\includegraphics[width=2.3in]{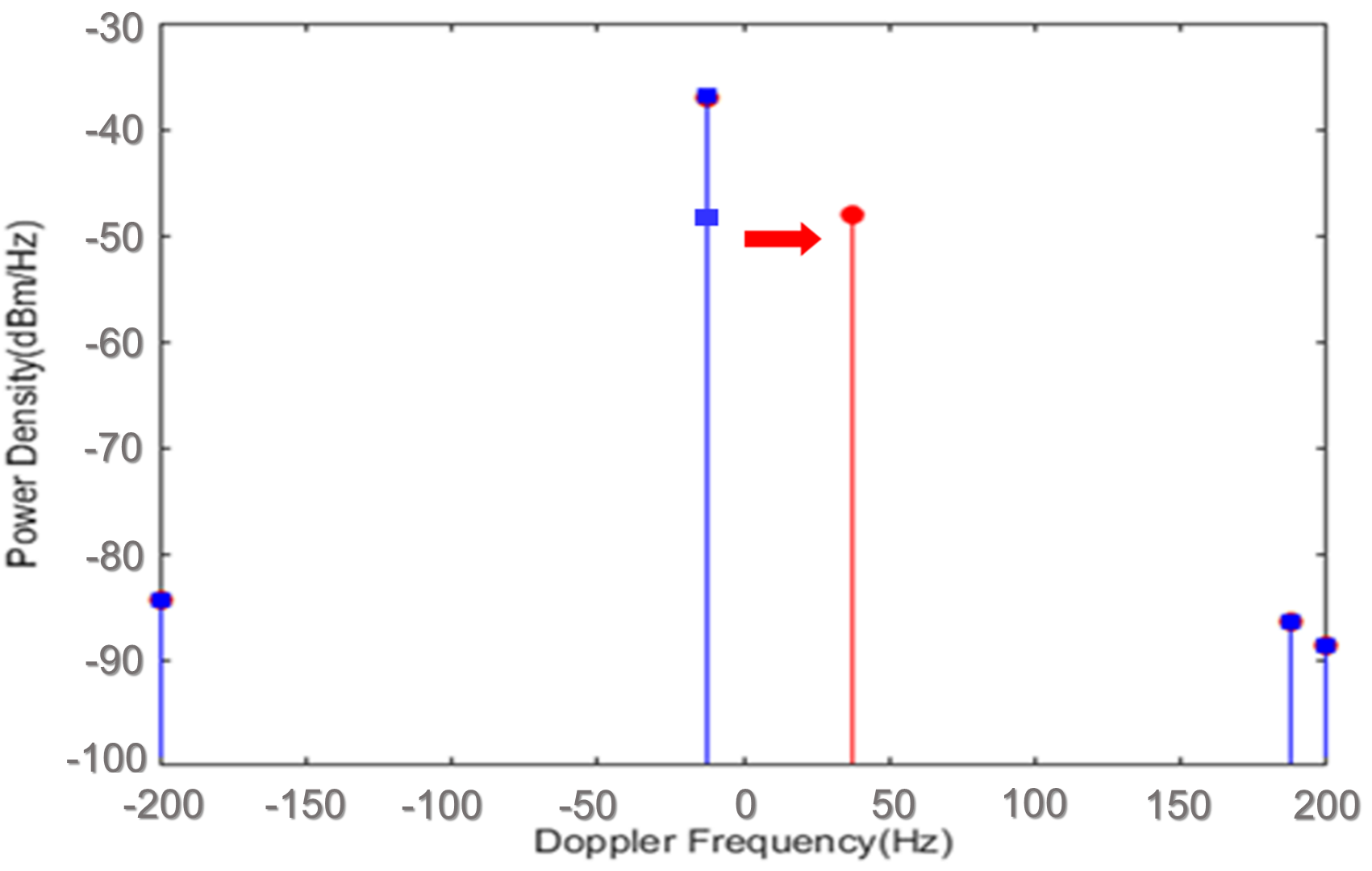}
	\caption{PDfP for a snapshot of the V2V scenario with reflection from bus side (see Fig. \ref{bus_aside}). Here TX and RX are moving in the same direction at 28 km/ and 30 km/h, respectively, while bus is moving at 30 km/h in the opposite direction. Bus moving straight without (blue) and with (red) rotation.}
	\label{bus_rotation}
\end{figure}

We consider now the case of TX and RX moving in the same direction with constant speed at 28 km/h and 30 km/h, respectively, while the bus is moving in the opposite direction at 30 km/h. We consider a time instant when a reflection on the bus side wall is present, and the bus deviates from its straight path toward the center lane, by rotating counterclockwise with an instantaneous angular velocity $\omega=\pi/6$ [rad/s] (see Fig. \ref{bus_aside}). A comparison of the PDfP is shown in Fig. \ref{bus_rotation} for the case of the bus driving straight (blue) and the bus swerving toward the center lane (red), respectively.
In the first case, the LoS ray and the ray reflected from the bus side are overlapped in the figure with Doppler shift close to zero, as both are almost perpendicular to the TX's, RX's and reflection point's velocities.
In the latter case the Doppler's shift of bus reflection becomes positive (37 Hz) due to the reflection point's movement toward the central lane. \par 
This behaviour is important from the applications point of view: when Doppler frequency of a major multipath component abruptly changes, this could indicate a potentially dangerous situation, e.g. a vehicle swerving from its lane. 

\subsection{Case Study 2 : Comparison with measurements in a street intersection}

\begin{figure}[h!]
    \centering
    \includegraphics[width=2.1in]{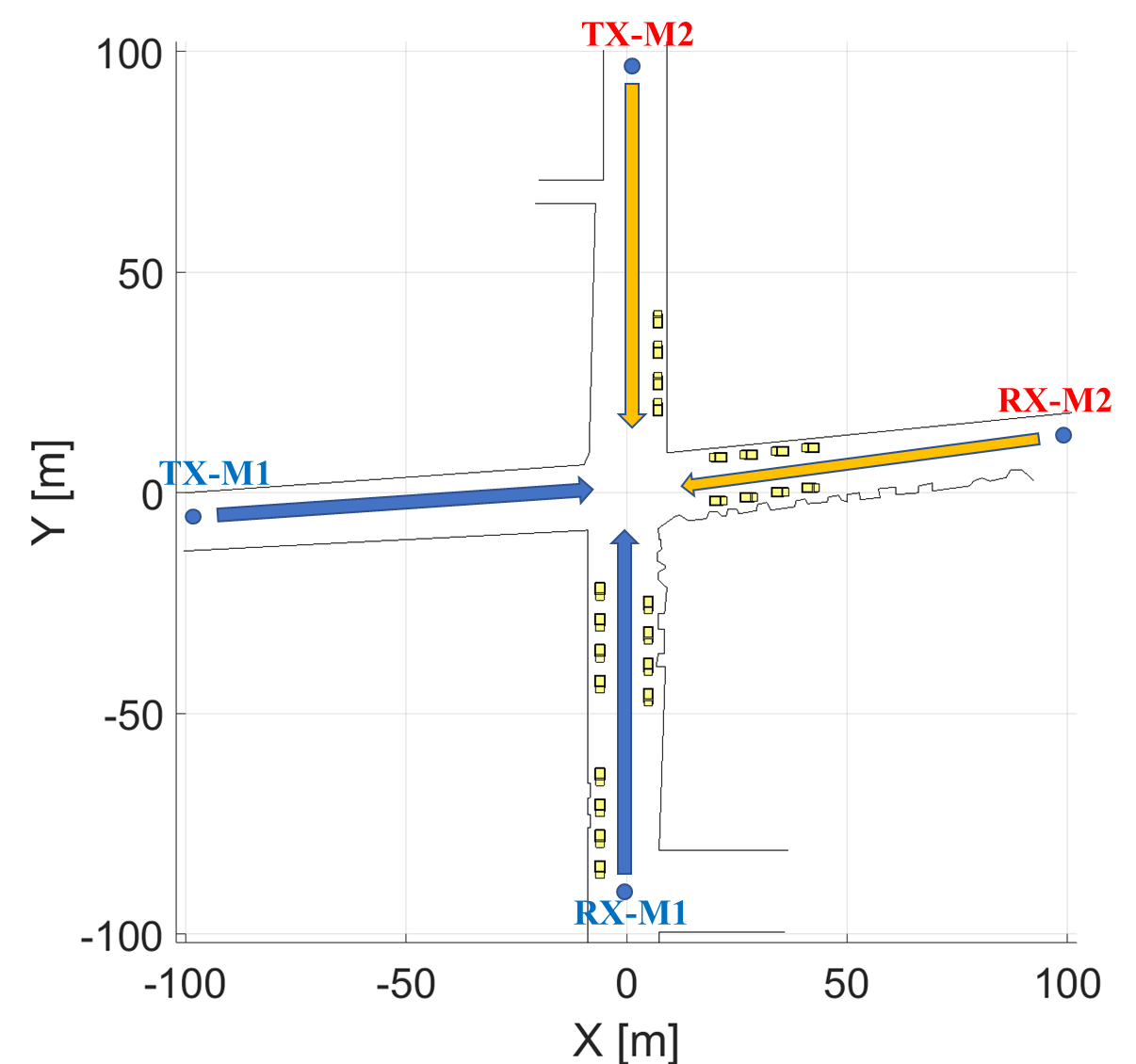}
    \caption{Intersection scenario in the city of Lund, Sweden}
    \label{lund}
\end{figure}

In this section we perform a validation of DRT simulations results for a V2V scenario by means of a comparison with measured channel data \cite{abbas} and conventional, multiple-snapshot RT simulation using the same 3D RT algorithm used in our DRT model for initial simulation \cite{fuschini2015}. Both RT and DRT are performed with a maximum of 2 specular reflections, 1 diffraction and 1 diffuse scattering. \par
The measurements described in \cite{abbas} were performed using the RUSK LUND channel sounder at a carrier frequency of 5.9 GHz with a bandwidth of 240 MHz and TX power of 1 W. The antenna modules, identical for Tx and Rx, consist of 4 identical arrays mounted at 90° from each other and properly driven in order to simulate a quasi omnidirectional antenna. Two hatchback cars were used to carry TX and RX and the antenna modules were mounted on the roof-top \cite{paier2010}.  
Based on these information, we used  omni-directional half-wavelength dipole antennas for all RT and DRT simulations. 

\subsubsection{Power delay profile}
The scenario is a narrow urban intersection as shown in Fig. \ref{lund}. The measurements are divided into two parts: $M1)$ when TX and RX are driving from the streets $TX-M1$ and $RX-M1$, and $M2)$ when TX and RX are driving from $TX-M2$ and $RX-M2$, respectively, towards the intersection with a speed of approximately $10$ m/s. The RT simulations are performed using the classical approach based on snapshots repeated every $\Delta t=10$ms. The overall time-span is $t_S=10$s. Since the scenario evolution within $t_S$ includes a transition between NLoS and LoS then certainly $t_S>T_C$. Therefore DRT simulation is divided into several $T_C$, namely into 25 and 20\,$T_C$s for $M1$ and $M2$, respectively, with larger $T_C$s at the beginning of the routes and shorter ones close to the NLoS/LoS transition. Such a subdivision was optimized after a short trial-and-error procedure in order for DRT results to become very similar to RT results. During each $T_C$ the acceleration of the terminals is considered constant. 
In the first seconds, TX and RX are far from the street intersection and LOS between them is blocked by the buildings situated at the corners for both scenarios. In this period of time, we used only 2\,$T_C$. In the following time span which stretches till the appearance of LOS components (C) and (F) at approximately $6.8$s and $7.9$s in $M1$ and $M2$, respectively, new contributions appear that make the multipath structure change rapidly. At this stage the number of $T_C$ need to be increased and their duration reduced. In the last $0.4$s leading to the LOS, 4\,$T_C$ with a duration of $100$ms, are used in both scenarios. By doing so, we were able to capture most of the contributions in the time span. A further discussion on the criteria for the choice of $T_C$ is provided in the next sub-section. \par 

\begin{figure*} [t]
    \captionsetup[subfigure]{font=scriptsize,labelfont=scriptsize}
    \centering
    \subfloat[Measured \cite{abbas} \label{measure_M1}]{\includegraphics[width=2.85in]{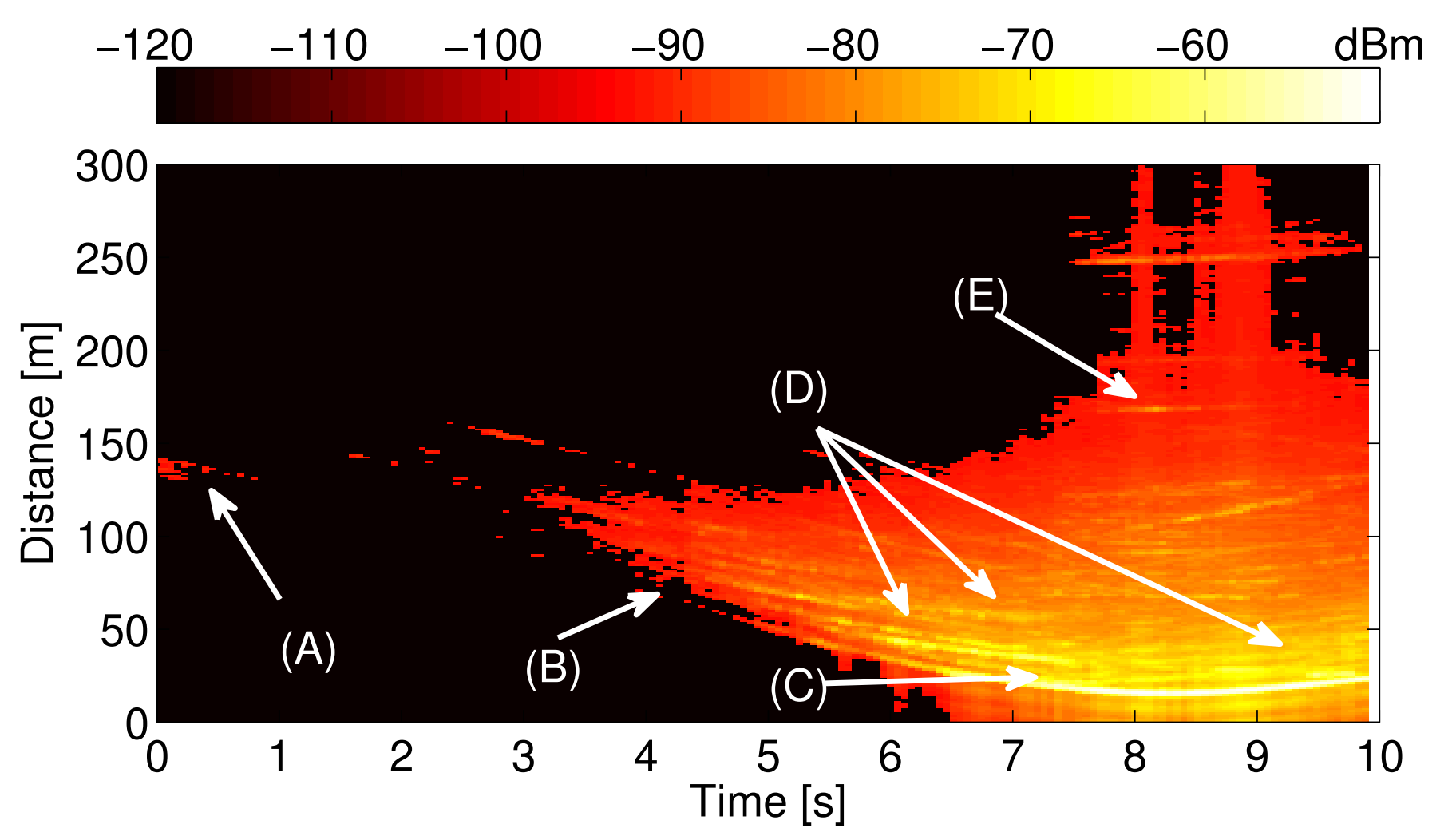}}
    \quad \quad \quad \quad
    \subfloat[Measured \cite{abbas}\label{measure_M2}]{\includegraphics[width=2.8in]{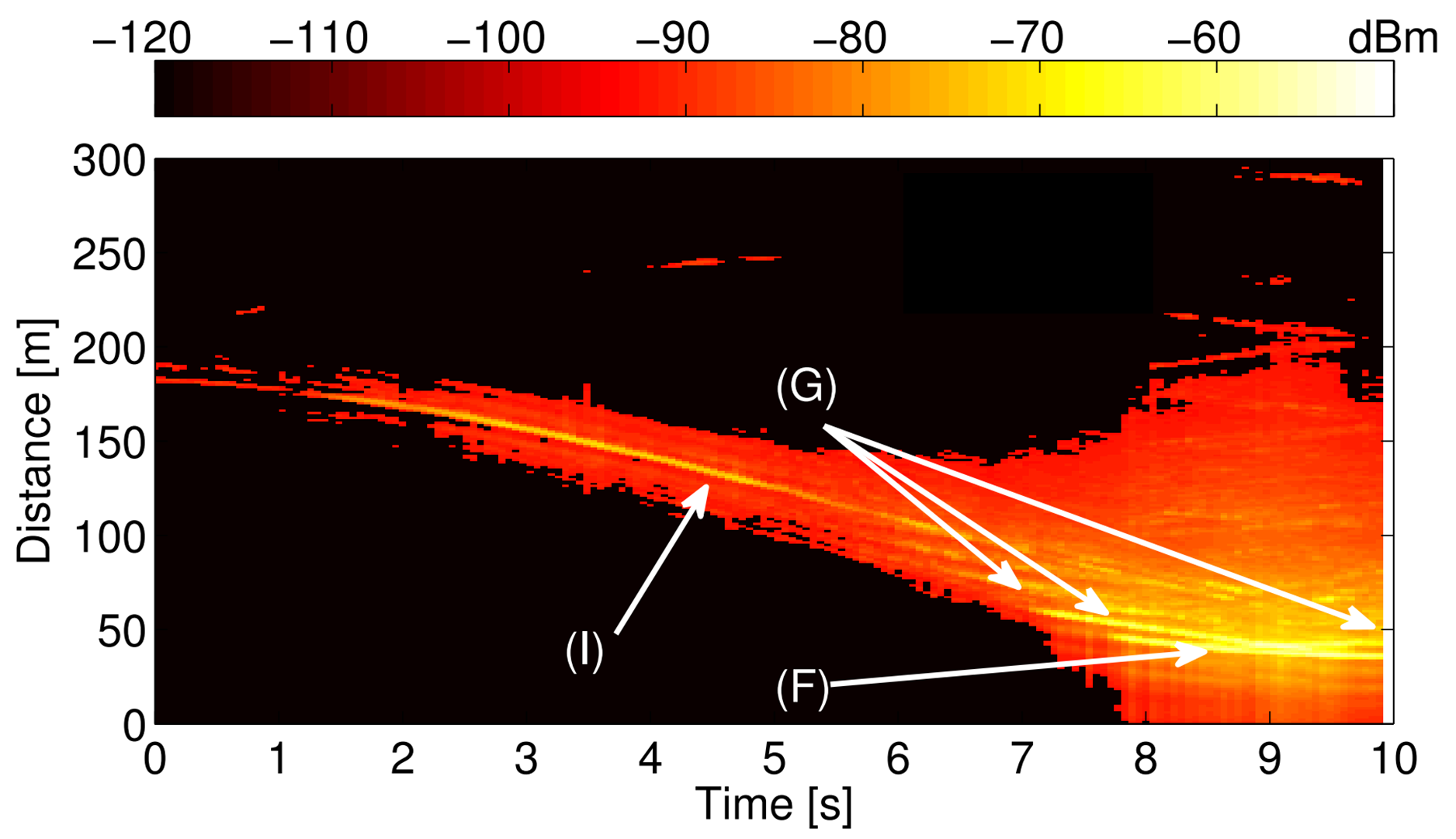}}
    \\
    \subfloat[RT simulation\label{RT_M1}]{\includegraphics[width=2.8in]{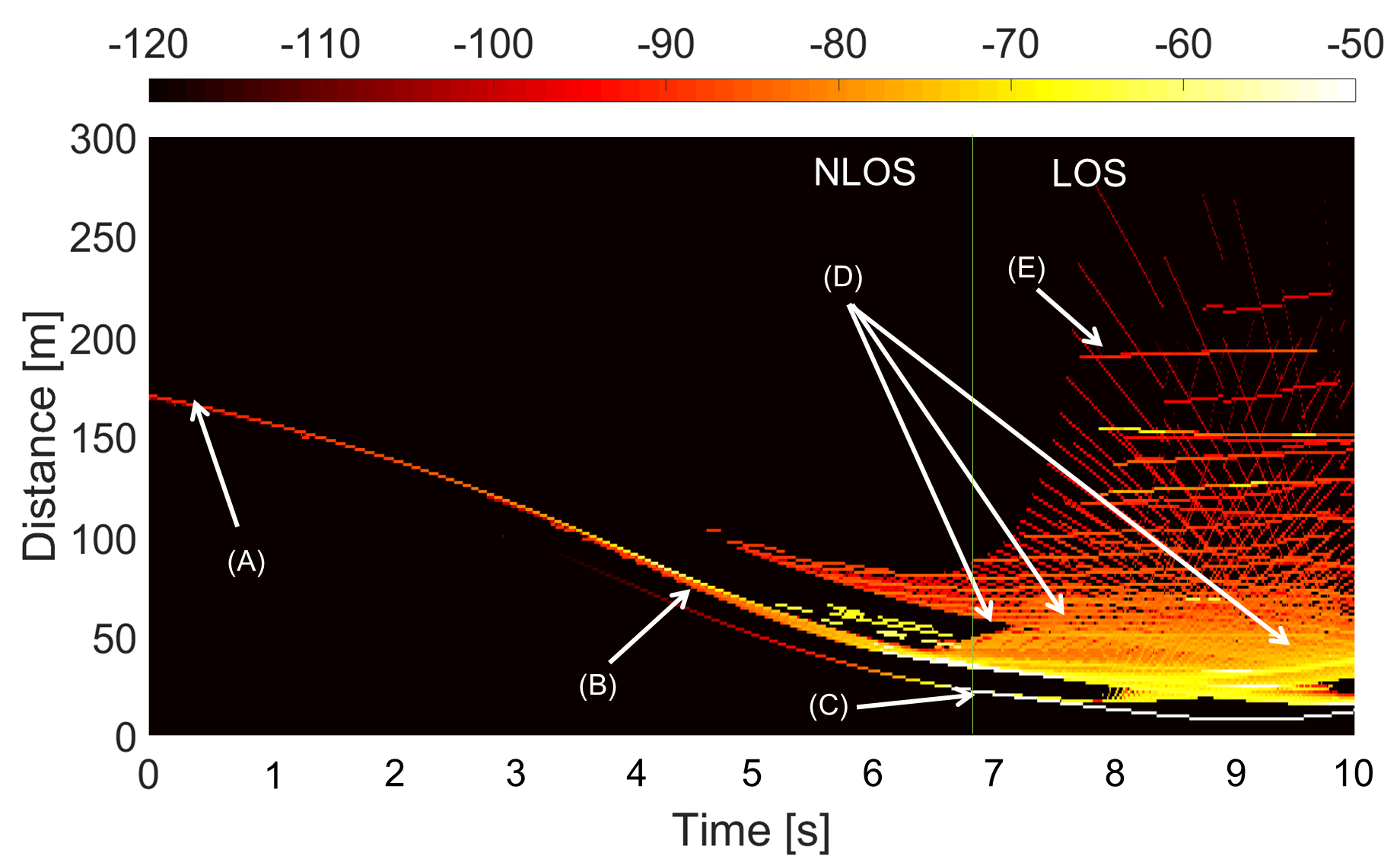}}
    \quad \quad \quad \quad
    \subfloat[RT simulation\label{RT_M2}]{\includegraphics[width=2.8in]{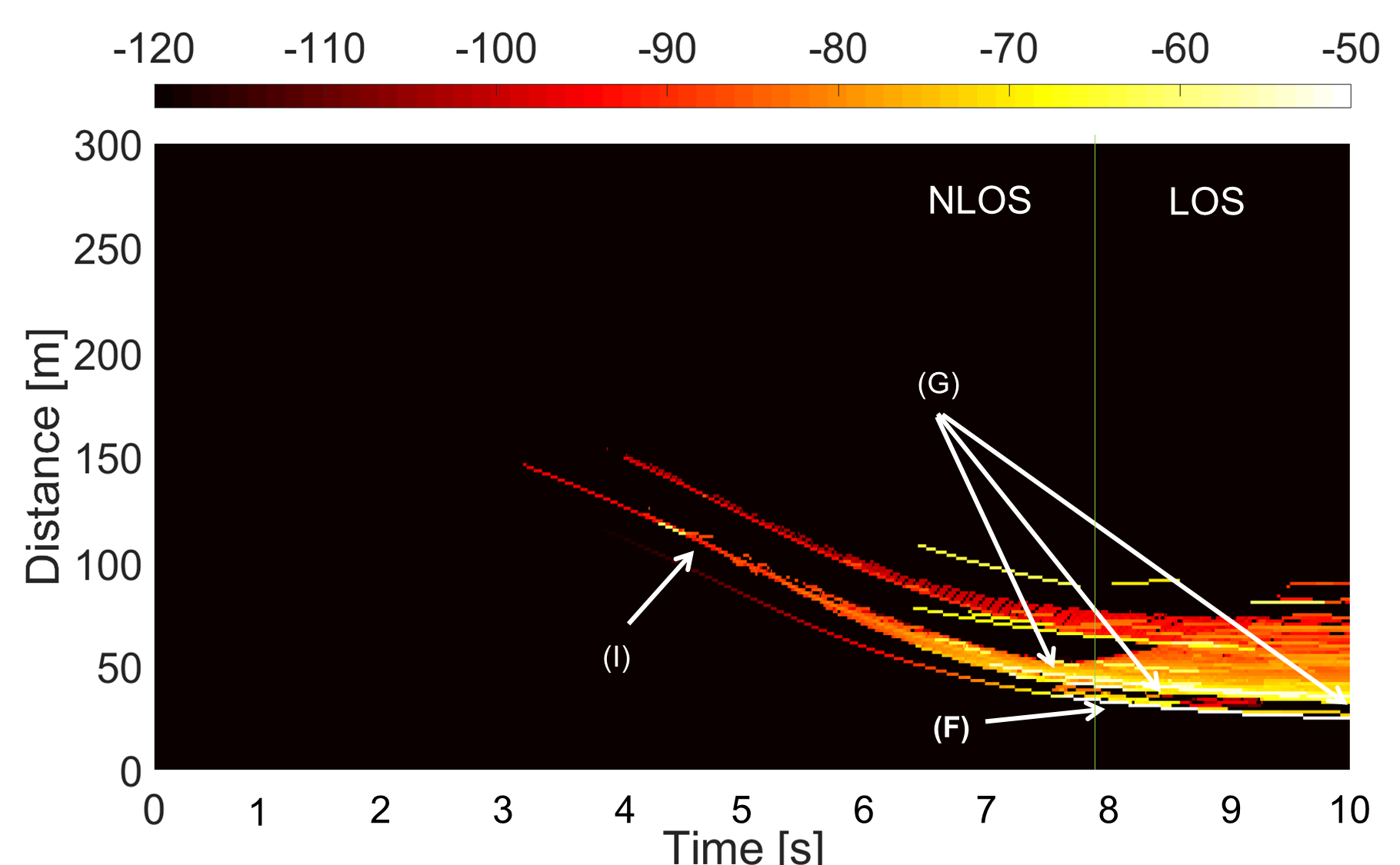}}
    \\
    \subfloat[DRT simulation\label{DRT_M1}]{\includegraphics[width=2.85in]{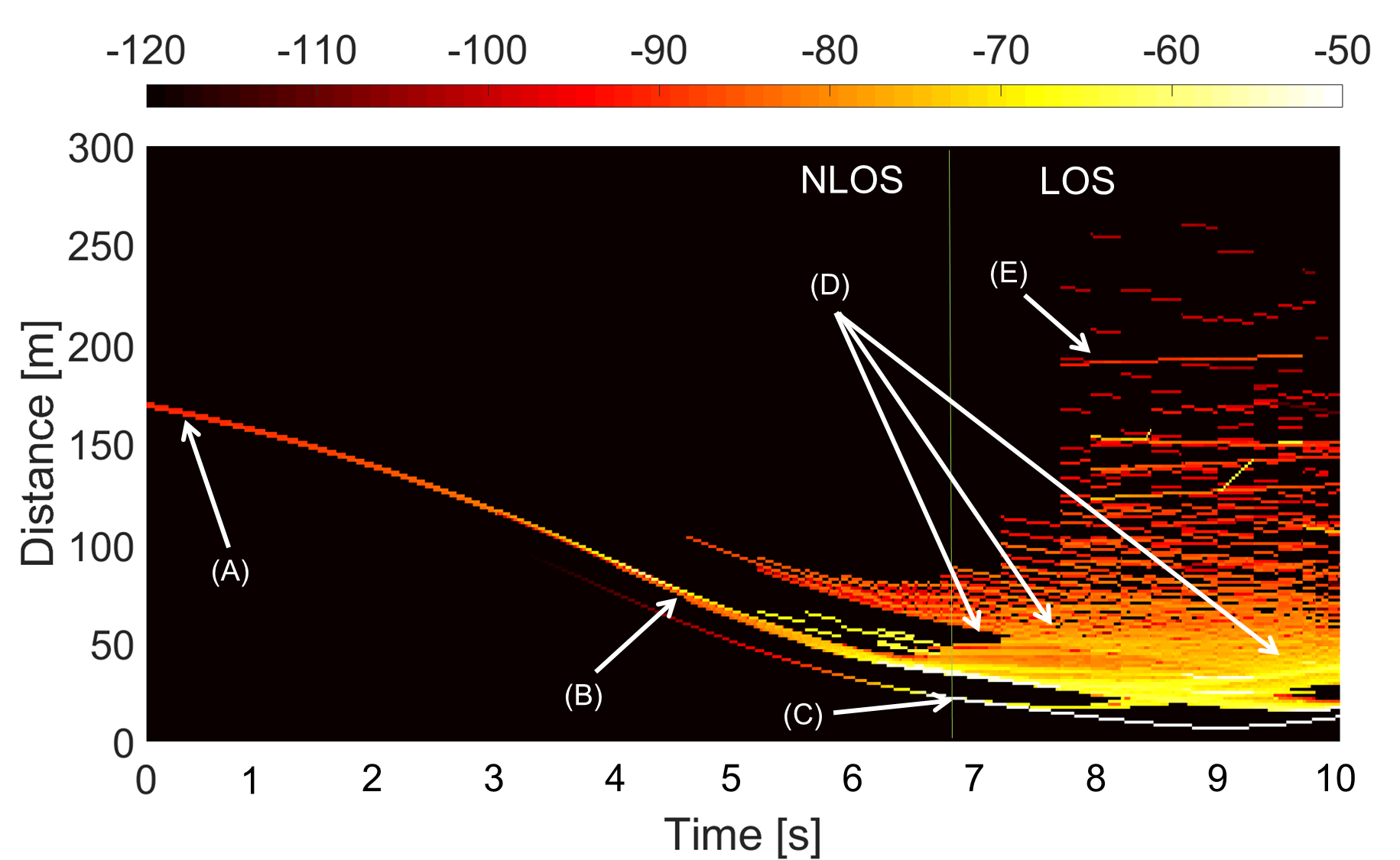}}
    \quad \quad \quad \quad
    \subfloat[DRT simulation\label{DRT_M2}]{\includegraphics[width=2.8in]{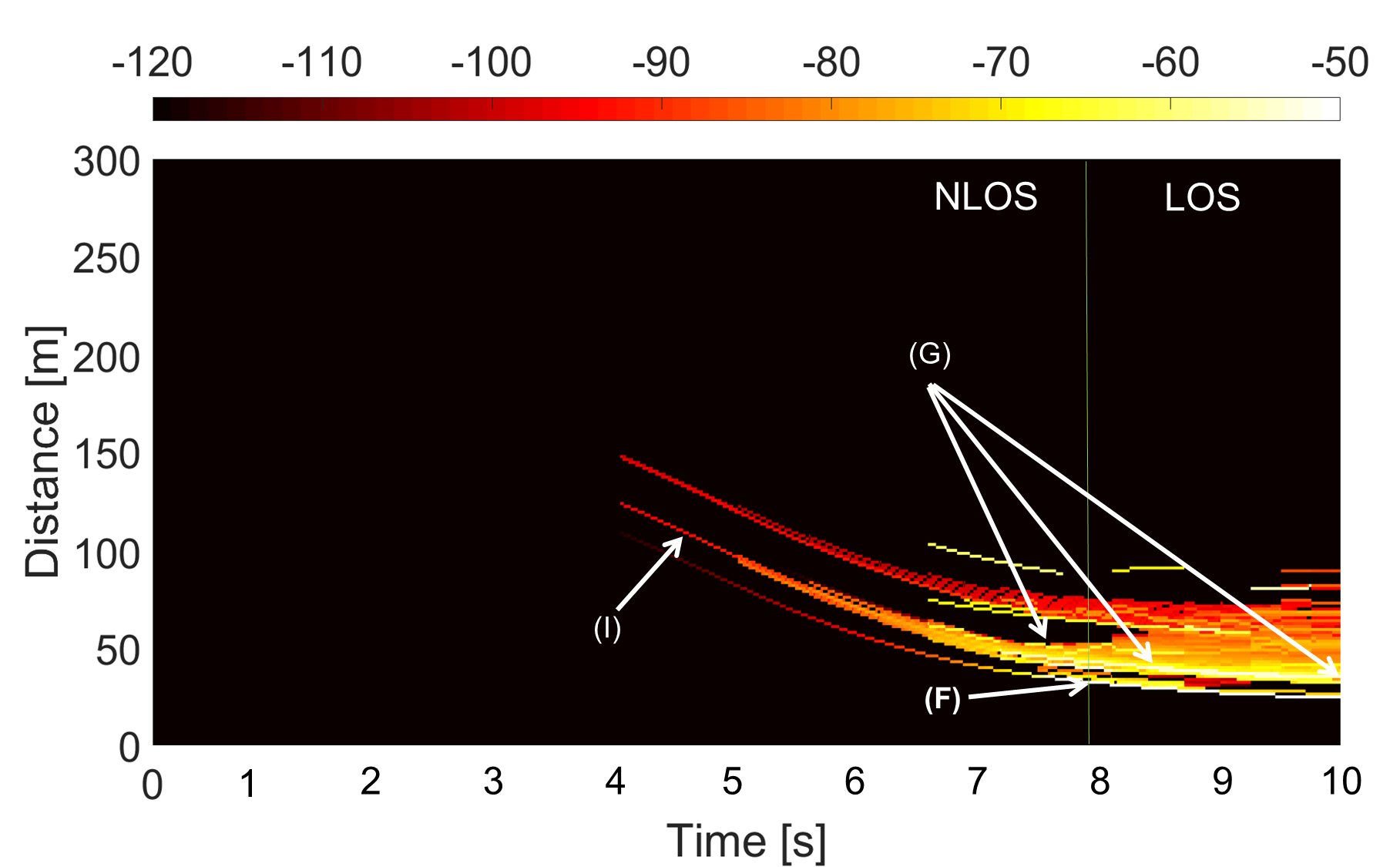}}
    \caption{Power Delay Profile obtained from measurements (a) and (b), from RT simulations (c) and (d) and from DRT prediction (e) and (f)}
    \label{PDP}
\end{figure*}

The measured, RT- and DRT-simulated power delay profiles (PDP) are depicted in Fig. \ref{measure_M1}, \ref{RT_M1}, \ref{DRT_M1} for $M1$ and Fig. \ref{measure_M2}, \ref{RT_M2}, \ref{DRT_M2} for $M2$. In general, there is good agreement between simulations and measurements with several similar contributions in the RT/DRT simulation and in the measurements. The group of arrows (D), (G) and (B) in Fig. \ref{PDP} point at contributions that probably originated from nearby buildings. There is also good agreement between our simulations and RT simulations performed in \cite{abbas} although our simulations appears to capture some more contributions in the final 2 seconds of $t_S$, probably because we considered a larger variety of interactions, including diffuse scattering from building walls and cars parked along the streets. However, there are several contributions in the measured PDP that are missing in the simulations. One of the reasons could be the incomplete building database used in RT/DRT. For example some of the contributions like (E) and (I), in \cite{abbas} are said to be from a building which has a metallic structure on the walls, which is not present in the building database.  This shows the necessity of a very detailed scenario description in order to get a really good match between simulations and measurements. \par

\subsubsection{Impact of the choice of $T_C$}
\begin{figure*}[t]
    
    \captionsetup[subfigure]{font=scriptsize,labelfont=scriptsize}
    \centering
    \subfloat[\label{3Tc}]{\includegraphics[width=2.8in]{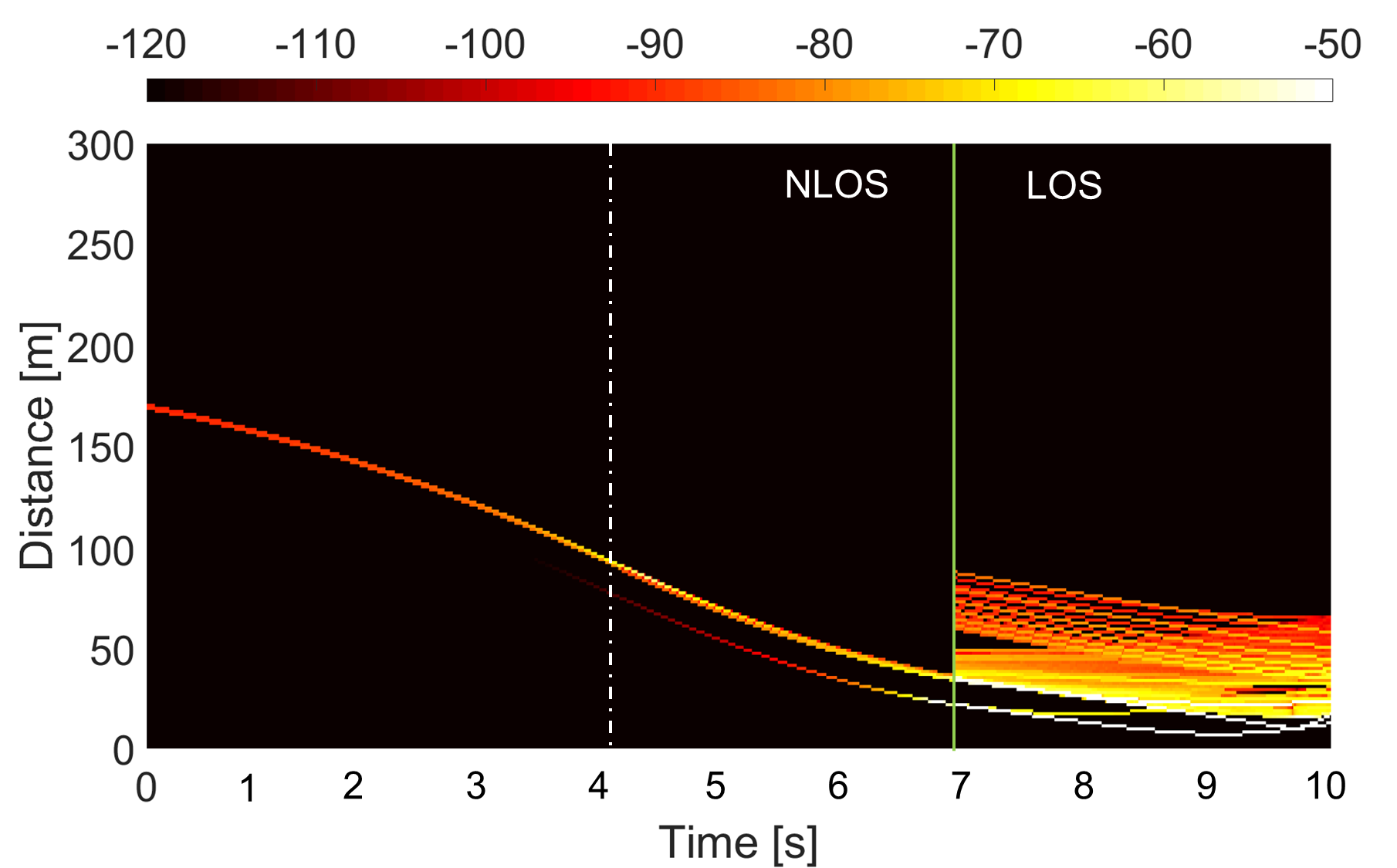}}
    \quad \quad \quad \quad
    \subfloat[\label{10Tc}]{\includegraphics[width=2.9in]{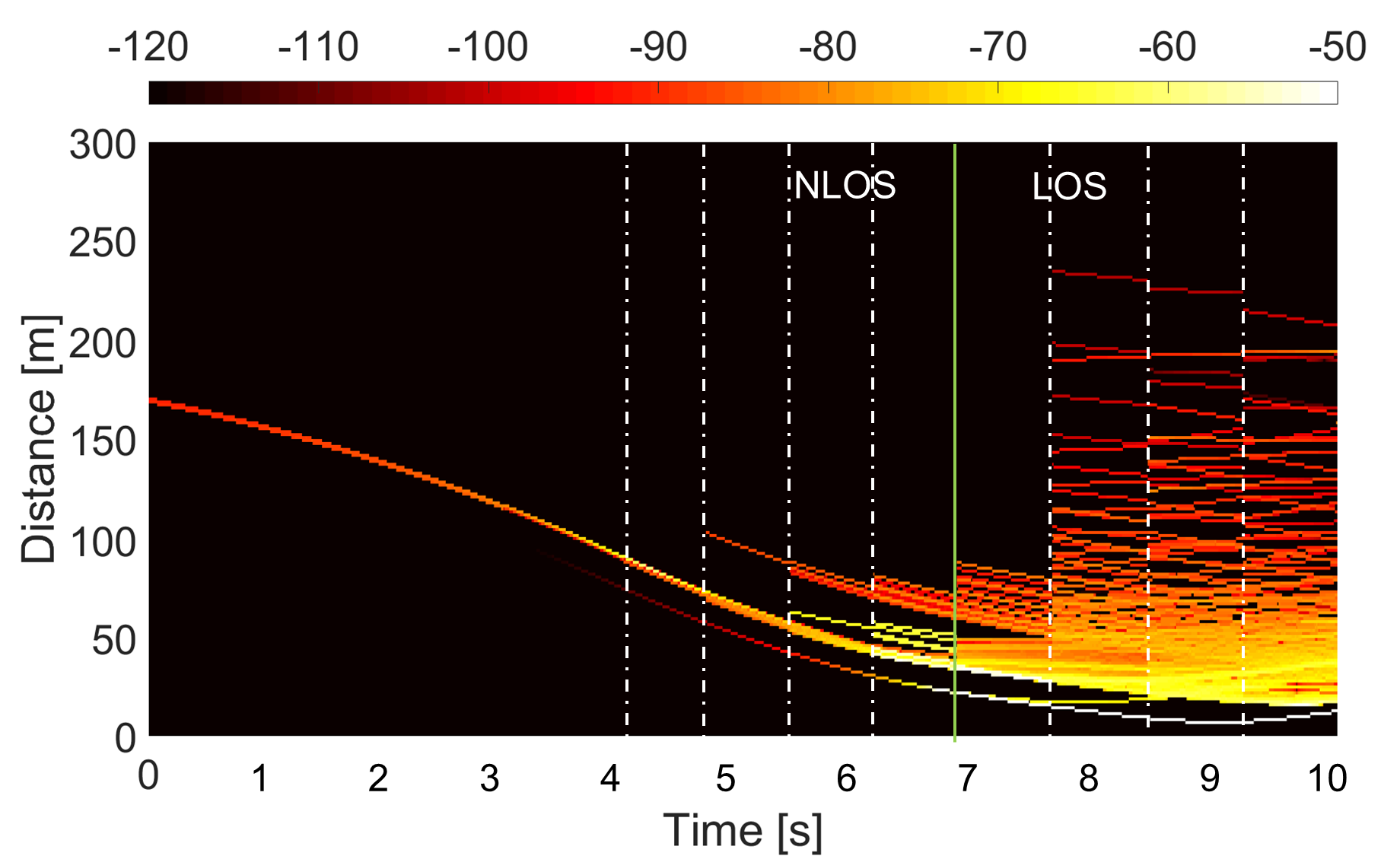}}
    \caption{PDP obtained from simulation with DRT. In (a) 3\,$T_C$ were used, in (b) 10\,$T_C$}
\end{figure*}

The number of $T_C$ and the right choice of them is very important in  DRT simulation, that has a strong impact on both accuracy of results and computation time. In Fig. \ref{3Tc}, DRT simulation PDP for $M1$ is presented with only 3\,$T_C$. It is evident that results are incomplete if compared to Fig. \ref{DRT_M1} because dynamic changes in the multipath structure are under-sampled in this case and several paths aree missed. In Fig. \ref{10Tc}, 10\,$T_C$ has been used. The PDP in this case is more complete and many new contributions can be seen. Nevertheless a larger number of $T_C$ close to the NLoS/LoS transition is needed in order to achieve the more complete results of Fig. \ref{DRT_M1}.  

\subsection{Computational Gain vs. traditional RT}
To compare the computation time of DRT vs. RT, we recorded computation times with the two approaches in $M1$ and $M2$ on the same Intel(R) Core(TM) i5-8265U CPU @1.60GHz 1.80 GHz platform. In DRT, one traditional RT run is included for multipath initialisation at the beginning of each $T_C$. To achieve the desired graphical resolution 1000 runs are performed with both DRT and RT, but a large fraction of them only requires the computation of analytical extrapolation formulas when using DRT, with a great computation time gain. \par
From Table \ref{time_gain}, we can observe a CPU-time  gain of about 48 times for the  $M2$ scenario while the computational gain is even higher in the $M2$ scenario because the number of $T_C$ is smaller.

\begin{table} [h!]
    \caption{Execution time for RT and DRT in M1 and M2}
    \label{time_gain}
    \centering
    \begin{tabular}{c|c|c|c}
         Scenario&\makecell{Standard RT\\(1000 snapshots)} &\makecell{Analytical approach\\(DRT)} &\makecell{Speed-up\\Factor} \\
         \hline
      M1 &4486 seconds &97.9 seconds  & 45.8x \\
      \hline
      M2 &4384 seconds &88.8 seconds  & 48.2x\\
      \hline
    \end{tabular}
\end{table}

\subsection{Future prospect: Predictive Radio Awareness}
\begin{figure}[h!]
    \centering
    \includegraphics[width=3.1in]{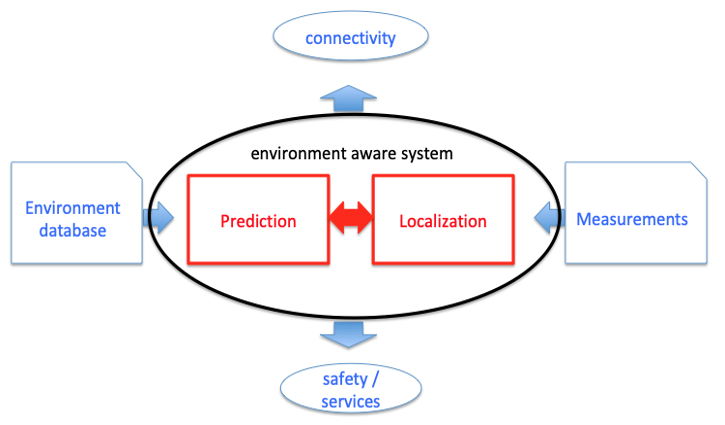}
    \caption{Scheme of an environment-aware system including both channel prediction and localization}
    \label{scheme_prediction}
\end{figure}

The  DRT approach to predict “ahead-of-time” (or anticipate) the environment and/or the radio channel characteristics in highly dynamic, industrial or vehicular applications and realize the so-called \emph{predictive radio awareness} or \emph{location aware communications} \cite{Kuerner2018,DiTaranto2014} is quite attractive. Exploiting such capabilities could be of paramount importance to guarantee reliable connectivity in critical application such as automated and connected driving and to foster interesting safety applications to detect dangerous situations in advance. Two kinds of applications are possible: 
\begin{enumerate}
    \item DRT-based radio channel anticipative prediction
    \item Environment configuration anticipative prediction
\end{enumerate}
In both cases, accurate localization of radio terminals and moving objects, which is likely to be available in future mobile radio systems \cite{Witrisal2016,Koivisto2017}, is a prerequisite. Accurate localization, together with the availability of a local environment database is used to build a dynamic environment database for the current time $T_0$.
In 1), DRT is used to extrapolate multipath characteristics, and therefore to estimate the Channel State Information for $t>T_0$. In 2) kinematics theory is used to extrapolate the environment configuration for $t>T_0$ and therefore detect possible hazards or collisions.  These applications will be fully addressed in follow-on studies.
It is worth noting that, in turn, the availability of techniques 1) and 2)  would be of great help to realize multipath-exploiting localization techniques  \cite{Thomae2018}. Therefore, the two goals of anticipative channel prediction and localization might be achieved in synergy to realize an environment-aware system and enhance both connectivity and safety, as depicted in Fig. \ref{scheme_prediction}.

\section{Conclusions}
In the present work, a complete formulation of a full-3D Dynamic ray tracing (DRT) algorithm is presented, that includes multiple reflections, edge diffraction and diffuse scattering for the general case of moving object that can translate and rotate. The approach allows to predict the multipath evolution within a given "multipath lifetime" $T_C$ on the base of a single conventional ray tracing prediction of the current multipath geometry, assuming constant speeds or accelerations for moving objects, using analytical, and therefore fast, extrapolation formulas.
The model is validated vs. traditional, multi-snapshot ray tracing prediction in an ideal, reference case, and in a real-life case presented in the literature, where it is also compared to measurements in terms of Power-Doppler Profiles. Results show a good agreement with measurements and very good agreement with traditional ray tracing prediction if the considered time-span is properly subdivided into mutiple $T_C$. The proper value of $T_C$ for vehicular applications is shown to be of the order 100 ms in the vicinity of abrupt LoS to NLoS transitions, and of the order of seconds elsewhere. The corresponding computation time speed-up with respect to ray tracing is shown to be of the order of 50x when a prediction resolution of 10 ms over a time-span of 10 s (1000 snapshots) is required.
Finally, interesting real-time applications of DRT are presented, such as ahead-of-time (or anticipative) field prediction to help Channel State Information estimation.


\appendices
\section{Doppler Frequency Calculation}
One of the advantages of the DRT approach is the computation of Doppler information online in the algorithm with the aid of simple formulas. In such a way, there is no need to consider successive “snapshots” of the environment with slightly different displacements of the objects, and then to calculate the Doppler shifts with a “finite difference” computation method. \par 
	When TX and RX are both moving, the resulting apparent frequency $f^{'}$, including the Doppler frequency shift $f_D$ is computed for the LoS ray using the following equation \cite{ChenDoppler}: 
	\begin{equation}
	f^{'} = f_{0} + f_{D} = f_{0} \left (\frac{c-\overline{v}_{RX}\cdot \hat{k}}{c-\overline{v}_{TX}\cdot \hat{k}} \right) 
	\end{equation}
	where $f_0$ is the carrier frequency of the transmitted signal, $\hat{k}$ is the unit vector of the ray's direction from TX towards RX, and $\overline{v}_{TX}$, $\overline{v}_{RX}$ are the velocities of transmitter and receiver, respectively. \par
\begin{figure}[h!]
		\centering
		\includegraphics[width=3in]{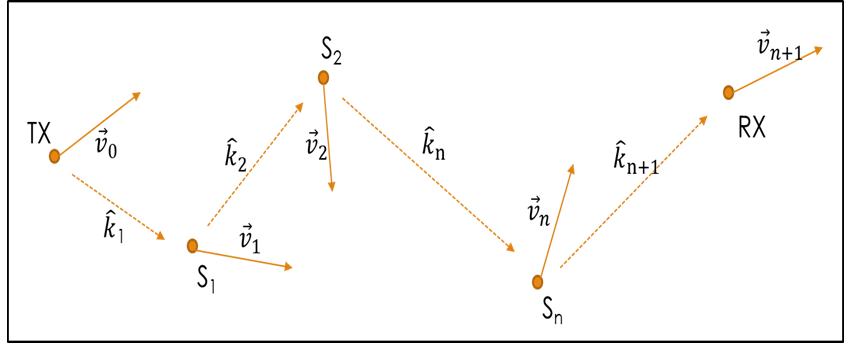}
		\caption{Representation of the multiple scatterers Doppler model}
		\label{doppler}
	\end{figure}

	This formula can be extended to rays with multiple bounces, where we have $n$ scattering points, each one moving with different speed (see Fig. \ref{doppler}): 
	\begin{equation}
	f^{'} = f_{0} + f_{D} = f_{0} \prod_{i=1}^{n+1} \left(\frac{c-\overline{v}_{i}\cdot \hat{k}_i}{c-\overline{v}_{i-1}\cdot \hat{k}_i}  \right)
	\label{doppler_freq}
	\end{equation}	
    where $\overline{v}_i$, $i=1,2,..,n$ is the velocity of the i-th interaction point, $\overline{v}_0=\overline{v}_{TX}$, and $\overline{v}_{n+1}=\overline{v}_{RX}$, respectively.\par
	Equation (\ref{doppler_freq}) assumes that the velocities of the interaction points on the objects are known. In the simplified case of small scattering objects that can be approximated as "point scatterers", no further processing is needed, and we can directly apply eq. (\ref{doppler_freq}). \par 
	Instead, in the case of large objects we need to compute the velocity of the interaction points on the object surface as explained in section III. 

\section{Reflection Point's Velocity and Acceleration Calculation}
	Given the reflection point position ($Q_R$), its velocity ($\overline{v}_{Q_{R}}$) can be determined by deriving eq. (\ref{ref_point}) with respect to time. The x-component of $\overline{v}_{Q_{R}}$ can be calculated as: 
	
	\begin{equation}
	\begin{gathered}
	v_{Q_{R,x}} = \frac{\partial x_{Q_R}}{\partial t} = \frac{\partial x_{Q_R}}{\partial x_{TX}}~\frac{\partial x_{TX}}{\partial t} + \frac{\partial x_{Q_R}}{\partial y_{TX}}~\frac{\partial y_{TX}}{\partial t} \\ 
	+ \frac{\partial x_{Q_R}}{\partial x_{RX}}~\frac{\partial x_{RX}}{\partial t} + \frac{\partial x_{Q_R}}{\partial y_{RX}}~\frac{\partial y_{RX}}{\partial t}  \\ 
	= f_{x_{TX}} + f_{y_{TX}} + f_{x_{RX}} + f_{y_{RX}}, 
	\end{gathered}
	\label{vQRx}
	\end{equation}
	with 
	\begin{equation*}
	\begin{gathered}
	    \begin{cases*}
	        f_{x_{TX}} = \frac{\partial x_{Q_R}}{\partial x_{TX}}~v_{TX,x} \\
	        f_{y_{TX}} = \frac{\partial x_{Q_R}}{\partial y_{TX}}~v_{TX,y} \\ 
	        f_{x_{RX}} = \frac{\partial x_{Q_R}}{\partial x_{RX}}~v_{RX,x} \\ 
	        f_{y_{RX}} = \frac{\partial x_{Q_R}}{\partial y_{RX}}~v_{RX,y}.
	    \end{cases*}
	\end{gathered}
   \end{equation*}
	
The partial derivatives in (\ref{vQRx}) can be easily computed by deriving $x_{Q_R}$ w.r.t. the $x$,$y$ coordinates of TX and RX: 
\begin{equation*}
	\begin{gathered}
	\begin{cases*}
	\frac{\partial x_{Q_R}}{\partial x_{TX}} = 1 - \frac{y_{TX}}{y_{TX}+y_{RX}}\\ 
	\frac{\partial x_{Q_R}}{\partial y_{TX}} = \frac{y_{RX} (x_{RX}-x_{TX})}{(y_{RX}+y_{TX})^2}\\
	\frac{\partial x_{Q_R}}{\partial x_{RX}} = \frac{y_{TX}}{y_{TX}+y_{RX}} \\ 
	\frac{\partial x_{Q_R}}{\partial y_{RX}} = \frac{-y_{RX} (x_{RX}-x_{TX})}{(y_{RX}+y_{TX})^2}
	\end{cases*}
	\end{gathered} 
\end{equation*}
and substituting these expressions in (\ref{vQRx}) we finally get $v_{Q_R,x}$. \par
By following the same method as above, we can obtain the z-component of the velocity, $v_{Q_R,z}$, by time deriving the z-coordinate of $Q_R$ ($z_{Q_R}$). \par 
A similar procedure is used to calculate $\overline{a}_{Q_{R}}$ by deriving $\overline{v}_{Q_{R}}$ with respect to time:
\begin{equation}
	\begin{gathered}
	a_{Q_R,x}=a_{Q_R,x}^{(1)}+a_{Q_R,x}^{(2)}+a_{Q_R,x}^{(3)}+a_{Q_R,x}^{(4)} \\
	= \frac{\partial}{\partial t} f_{x_{TX}}+\frac{\partial}{\partial t} f_{y_{TX}}+\frac{\partial}{\partial t} f_{x_{RX}}+\frac{\partial}{\partial t} f_{y_{RX}}
	\end{gathered}
	\label{a_QRx}
\end{equation}
For the sake of brevity, we report below only the computation of the first partial derivative in eq. (\ref{a_QRx}): 
\begin{equation}
	\begin{gathered}
	a_{Q_R,x}^{(1)} = \frac{\partial}{\partial t} f_{x_{TX}} = \frac{\partial}{\partial t} \frac{\partial x_{Q_R}}{\partial x_{TX}} \frac{\partial x_{TX}}{\partial t} + \frac{\partial x_{Q_R}}{\partial x_{TX}} \frac{v_{TX,x}}{\partial t} \\ 
	= \left(\frac{\partial^2 x_{Q_R}}{\partial x_{TX} \partial y_{TX}} \frac{\partial y_{TX}}{\partial t} + \frac{\partial^2 x_{Q_R}}{\partial x_{TX} \partial y_{RX}} \frac{\partial y_{RX}}{\partial t} \right) v_{TX,x} \\
	+ \frac{\partial x_{Q_R}}{\partial x_{TX}} a_{TX,x}\\ 
	= \left(\frac{\partial^2 x_{Q_R}}{\partial x_{TX} \partial y_{TX}} v_{TX,y}  + \frac{\partial^2 x_{Q_R}}{\partial x_{TX} \partial y_{RX}} v_{RX,y} \right) v_{TX,x} \\ 
	+ \frac{\partial x_{Q_R}}{\partial x_{TX}} a_{TX,x}
	\end{gathered}
	\label{aQRx1}
\end{equation}
	where the partial derivatives in (\ref{aQRx1}) are expressed by:
	\begin{equation*}
	\begin{gathered}
	\begin{cases*}
	\frac{\partial x_{Q_R}}{\partial x_{TX}} = 1 - \frac{y_{TX}}{y_{TX}+y_{RX}}\\ 
	\frac{\partial^2 x_{Q_R}}{\partial x_{TX} \partial y_{TX}} = \frac{-y_{RX}}{\left( y_{TX}+y_{RX}\right)^2}\\ 
	\frac{\partial^2 x_{Q_R}}{\partial x_{TX} \partial y_{RX}} = \frac{y_{TX}}{\left( y_{TX}+y_{RX}\right)^2}.
	\end{cases*}
	\end{gathered}
	\end{equation*}
	By substituting these expressions in (\ref{aQRx1}) we obtain $a_{Q_R,x}^{(1)}$. \par
	By repeating the same procedure for the remaining components of $a_{Q_R,x}$, we finally get: 
	\begin{equation}
			\begin{gathered}
			a_{Q_R,x} = \left(\frac{\partial^2 x_{Q_R}}{\partial x_{TX} \partial y_{TX}} v_{TX,y}  + \frac{\partial^2 x_{Q_R}}{\partial x_{TX} \partial y_{RX}} v_{RX,y} \right) v_{TX,x}  \\
			+\frac{\partial x_{Q_R}}{\partial x_{TX}} a_{TX,x} + \left(\frac{\partial^2 x_{Q_R}}{\partial y_{TX} \partial x_{TX}} v_{TX,x} + \frac{\partial^2 x_{Q_R}}{\partial y_{TX} \partial x_{RX}} v_{RX,x} \right) v_{TX,y} \\ 
			+ \left (\frac{\partial^2 x_{Q_R}}{\partial y_{TX}^2 } v_{TX,y} + \frac{\partial^2 x_{Q_R}}{\partial y_{TX} \partial y_{RX} } v_{RX,y}  \right) v_{TX,y} + \frac{\partial x_{Q_R}}{\partial y_{TX}} a_{TX,y} \\ 
			+ \left(\frac{\partial^2 x_{Q_R}}{\partial x_{RX} \partial y_{TX}} v_{TX,y} + \frac{\partial^2 x_{Q_R}}{\partial x_{RX} \partial y_{RX}} v_{RX,y} \right) v_{RX,x} + \frac{\partial x_{Q_R}}{\partial x_{RX}} a_{RX,y} \\ 
			+ \left( \frac{\partial^2 x_{Q_R}}{\partial y_{RX} \partial x_{TX}} v_{TX,x} +  \frac{\partial^2 x_{Q_R}}{\partial y_{RX} \partial x_{RX}} v_{RX,x}  \right) v_{RX,y} \\ 
			+ \left(  \frac{\partial^2 x_{Q_R}}{\partial y_{RX} \partial y_{TX}} v_{TX,y} +  \frac{\partial^2 x_{Q_R}}{\partial y_{RX}^2} v_{RX,x}\right) v_{RX,x} + \frac{\partial x_{Q_R}}{\partial y_{RX}} a_{RX,y}.
			\end{gathered}
    \end{equation}
	
	A similar approach can be adopted to compute the z-component of  $\overline{a}_{Q_{R}}$.

\section{Diffraction Point's Velocity Calculation}
Time derivation of (\ref{diff_point}), gives us the z-component of the diffraction point instantaneous velocity: 
	\begin{equation}
	\begin{gathered}
	v_{Q_{D,z}} = \frac{\partial z_{Q_D}}{\partial t} = \frac{\partial z_{RX}}{\partial t} + \frac{\partial }{\partial t} \left(\frac{d_{RX}}{d_{RX}+d_{TX}} (z_{TX}-z_{RX}) \right) \\
	= v_{RX,z} + \frac{d_{RX}}{d_{RX}+d_{TX}} \left(v_{TX,z}-v_{RX,z} \right)\\ 
	+ (z_{TX}-z_{RX}) \frac{\partial}{\partial t} \left(\frac{d_{RX}}{d_{RX}+d_{TX}} \right)
	\end{gathered}
	\label{diff_point_velo}
	\end{equation}
	where \\
	\begin{equation}
	\begin{multlined}
	\frac{\partial}{\partial t} \left(\frac{d_{RX}}{d_{TX}+d_{RX}} \right)= \\[2ex]
	=\frac{\frac{\partial d_{RX}}{\partial t} (d_{RX}+d_{TX})-\left(\frac{\partial d_{RX}}{\partial t} + \frac{\partial d_{TX}}{\partial t} \right) d_{RX}}{(d_{TX}+d_{RX})^2}
	\end{multlined}
	\label{diff_point_velo2}
	\end{equation}
    \par 
	The derivative of $d_{RX}$ with respect to time can be computed by applying the derivative chain rule to eq. (\ref{dist2D_TXRX}):
	\begin{equation}
	\begin{gathered}
	\frac{\partial d_{RX}}{\partial t} = \frac{\partial d_{RX}}{\partial x_{RX}} \frac{\partial x_{RX}}{\partial t} + \frac{\partial d_{RX}}{\partial y_{RX}} \frac{\partial y_{RX}}{\partial t} \\ 
	= \frac{1}{d_{RX}}\left[ \left(x_{RX}-x_{Q_D}\right) v_{RX,x}+  \left(y_{RX}-y_{Q_D}\right) v_{RX,y}\right].
	\end{gathered}
	\label{diff_point_velo3}
	\end{equation}
	In similar way the derivative of $d_{TX}$ can be calculated: 
	\begin{equation}
	\frac{\partial d_{TX}}{\partial t} = \frac{1}{d_{TX}}\left[ \left(x_{TX}-x_{Q_D}\right) v_{TX,x}+  \left(y_{TX}-y_{Q_D}\right) v_{TX,y}\right]
	\label{diff_point_velo4}
	\end{equation}
By substituting (\ref{diff_point_velo3}) and (\ref{diff_point_velo4}) into (\ref{diff_point_velo2}) and (\ref{diff_point_velo}), we finally get $v_{Q_D,x}$.\par
The diffraction point's acceleration ($a_{Q_{D,z}}$) can be computed in a similar way by time deriving (\ref{diff_point_velo}).

\section*{}


\begin{thebibliography}{}

\bibliographystyle{IEEEtran}

\bibitem{iskander}
    Z. Yun and M. F. Iskander, "Ray Tracing for Radio Propagation Modeling: Principles and Applications," \textit{IEEE Access}, Vol. 3, pp. 1089-1100, 2015.
\bibitem{fuschini2019}
    F. Fuschini, M. Zoli, E.M. Vitucci, M. Barbiroli and V. Degli-Esposti, “A Study on Mm-wave Multi-User Directional Beamforming Based on Measurements and Ray Tracing Simulations,” \textit{IEEE Trans. Antennas Propag.}, Vol 67, No. 4, pp 2633-2644, Apr. 2019.
\bibitem{Bhat2018}
    A. Bhat, S. Aoki and R. Rajkumar, "Tools and Methodologies for Autonomous Driving Systems," \textit{Proc. IEEE}, Vol. 106, No. 9, pp. 1700-1716, Sep. 2018.
\bibitem{boban1}
    R. He et al., "Propagation Channels of 5G Millimeter-Wave Vehicle-to-Vehicle Communications: Recent Advances and Future Challenges," \textit{IEEE Veh. Technol. Mag.}, Vol. 15, No. 1, pp. 16-26, Mar 2020.
\bibitem{Xing2020}
    Y. Xing, C. Lv and D. Cao, "Personalized Vehicle Trajectory Prediction Based on Joint Time-Series Modeling for Connected Vehicles," \textit{IEEE Trans. Veh. Technol.}, Vol. 69, No. 2, pp. 1341-1352, Feb. 2020.
\bibitem{VDE2021}
    V. Degli-Esposti, “Ray tracing: techniques, applications and prospect,” in Proc. of 2020 International Symposium on Antennas and Propagation, Osaka, Japan, pp. 1-2, January 25-28, 2021.
\bibitem{he2019}
    D. He, B. Ai, K. Guan, L. Wang, Z. Zhong and T. Kürner, "The Design and Applications of High-Performance Ray-Tracing Simulation Platform for 5G and Beyond Wireless Communications: A Tutorial," \textit{IEEE Commun. Surv. Tutor.}, Vol. 21, No. 1, pp. 10-27, Firstquarter 2019.
\bibitem{hussain2019}
    S. Hussain and C. Brennan, “Efficient preprocessed ray tracing for 5G mobile transmitter scenarios in urban microcellular environments,” \textit{IEEE Trans. Antennas Propag.}, Vol. 67, No. 5, pp. 3323–3333, May 2019.
\bibitem{nuckelt2015}
    J. Nuckelt, M. Schack, and T. Kürner, “Geometry-based path interpolation for rapid ray-optical modeling of vehicular channels,” in Proc. of 9th Eur. Conference on Antennas and Propagation (EUCAP), 2015, pp. 1–5, Lisbon, Portugal, Apr. 12-17, 2015.
\bibitem{azpilicueta}    
    L. Azpilicueta, C. Vargas-Rosales and F. Falcone, "Intelligent Vehicle Communication: Deterministic Propagation Prediction in Transportation Systems," \textit{IEEE Veh. Technol. Mag.}, Vol. 11, No. 3, pp. 29-37, Sep. 2016.
\bibitem{bilibashi}
    D. Bilibashi, E. M. Vitucci and V. Degli-Esposti, "Dynamic Ray Tracing: Introduction and Concept," in Proc. of 2020 14th European Conference on Antennas and Propagation (EuCAP), 2020, pp. 1-5.
\bibitem{bilibashi2}
    D. Bilibashi, E. M. Vitucci and V. Degli-Esposti, "Dynamic Ray Tracing: A 3D Formulation," in Proc. of 2020 International Symposium on Antennas and Propagation (ISAP), 2021, pp. 279-280.
\bibitem{qua}
    F. Quatresooz, S. Demey and C. Oestges, ”Tracking of Interaction Points for Improved Dynamic Ray Tracing,” \textit{IEEE Trans. Veh. Technol.}, Vol. 70, No. 7, pp. 6291-6301, Jul. 2021.
\bibitem{Martin1968}
    G. H. Martin, Kinematics and Dynamics of Machines, McGraw-Hill, 1968.
\bibitem{taylor2005}
    J. R. Taylor, Classical mechanics, University Science Books, 2005.
\bibitem{fuschini2015}
    F. Fuschini, E. M. Vitucci, M. Barbiroli, G. Falciasecca and V. Degli-Esposti, "Ray tracing propagation modeling for future small-cell and indoor applications: A review of current techniques," \textit {Radio Science}, Vol. 50, No. 6, pp. 469-485, Jun. 2015.
\bibitem{vitucci2019}
    E. M. Vitucci, J. Chen, V. Degli-Esposti, J. S. Lu, H. L. Bertoni, X. Yin, "Analyzing Radio Scattering Caused by Various Building Elements Using Millimeter-Wave Scale Model Measurements and Ray Tracing," \textit{IEEE Trans. Antennas Propag.}, Vol. 67, No. 1, pp. 665-669, Jan. 2019.
\bibitem{UTD} 
    R.G. Kouyoumjian, P. H. Pathak, "A uniform geometrical theory of diffraction for an edge in a perfectly conducting surface", \textit{Proc. IEEE}, vol. 62, No. 11, pp. 1448-1461, Nov. 1974. 
\bibitem{Keller} 
    J. B. Keller, “Geometrical theory of diffraction,” \textit{J. Opt. Soc. Amer.}, vol. 52, pp. 116-130, 1962.
\bibitem{abbas}
    T. Abbas, J. Nuckelt, T. Kürner, T. Zemen, C. F. Mecklenbräuker and F. Tufvesson, "Simulation and Measurement-Based Vehicle-to-Vehicle Channel Characterization: Accuracy and Constraint Analysis," \textit{IEEE Trans. Antennas Propag.}, Vol. 63, No. 7, pp. 3208-3218, Jul. 2015.
\bibitem{paier2010}
    A. Paier, L. Bernadó, J. Karedal, O. Klemp and A. Kwoczek, "Overview of Vehicle-to-Vehicle Radio Channel Measurements for Collision Avoidance Applications," \textit{IEEE 71st Vehicular Technology Conference,} 2010, pp. 1-5.
\bibitem{Kuerner2018}
    R. Alieiev, T. Hehn, A. Kwoczek and T. Kürner, “Predictive Communication and its Application to Vehicular Environments: Doppler-Shift Compensation,” \textit{IEEE Trans. Veh. Technol.}, Vol. 67, No. 8, pp. 7380-7393, Aug. 2018.
\bibitem{DiTaranto2014}
    R. Di Taranto, S. Muppirisetty, R. Raulefs, D.T.M. Slock, T. Svensson, H. Wymeersch, “Location-Aware Communications for 5G Networks”, IEEE Signal Proc. Magazine, vol. 31, no. 6, pp. 102-112, November 2014
\bibitem{Witrisal2016}
    K. Witrisal, P. Meissner, E. Leitinger, Y. Shen, C. Gustafson, F. Tufvesson, K. Haneda, D. Dardari, A. Molisch, A. Conti, and M. Z. Win, “High-accuracy localization for assisted living: 5G systems will turn multipath channels from foe to friend,” \textit{IEEE Signal Process. Mag.}, vol. 33, no. 2, pp. 59–70, Mar. 2016.
\bibitem{Koivisto2017}
    M. Koivisto, A. Hakkarainen, M. Costa, P.Kela, K. Leppanen, M. Valkama, “High-Efficiency Device Positioning and Location Aware Communications in Dense 5G Networks”, \textit{IEEE Comm. Mag.}, vol. 55, no. 8, pp. 188-195, Aug. 2017.
\bibitem{Thomae2018}
    M. N. de Sousa, R. S. Thomä,  “Enhancement of Localization Systems in NLOS Urban Scenario with Multipath Ray Tracing Fingerprints and Machine Learning,” \textit{Sensors}, Vol. 18, No. 11, 4073, 2018.
\bibitem{ChenDoppler} 
    V. C. Chen, The Micro-Doppler Effect in Radar, 2nd Edition, Artech House, 2019.









    

    

    








    
\end{thebibliography}
\end{document}